\documentclass[
aps,
,amsmath,amssymb,amsfonts
,superscriptaddress
,floatfix
,longbibliography
,twocolumn,10pt
]{revtex4-2}

\usepackage[T1]{fontenc}
\usepackage[utf8]{inputenc}
\usepackage{times}
\usepackage[low-sup]{subdepth}
\usepackage{wasysym}
\usepackage[usenames, dvipsnames]{color}
\usepackage[dvipsnames,svgnames,x11names,hyperref]{xcolor}

\usepackage{graphicx}

\definecolor{diffcolor}{RGB}{175,31,36} % Ironman
\usepackage[
  unicode=true,
  pdfusetitle,
  bookmarks=false,
  colorlinks=true,
  linkcolor=magenta,
  urlcolor=magenta,
  citecolor=blue,
  pdfencoding=auto
]{hyperref}

\newcommand{\subfigimg}[3][,]{%
  \setbox1=\hbox{\includegraphics[#1]{#3}}% Store image in box
  \leavevmode\rlap{\usebox1}% Print image
  \rlap{\hspace*{10pt}\raisebox{\dimexpr\ht1-1.1\baselineskip}{#2}}% Print label
  \phantom{\usebox1}% Insert appropriate spcing
}

\begin{document}

\title{Many-body energy invariant for $T$-linear resistivity}

\author{Aavishkar A. Patel}
\email{aavishkarpatel@berkeley.edu}
\affiliation{Department of Physics, University of California, Berkeley, CA 94720, USA}
\author{Hitesh J. Changlani}
\email{hchanglani@fsu.edu}
\affiliation{Department of Physics, Florida State University, Tallahassee, Florida 32306, USA}
\affiliation{National High Magnetic Field Laboratory, Tallahassee, Florida 32310, USA} 

\begin{abstract}
The description of dynamics of strongly correlated quantum matter is a challenge, particularly in physical situations where a quasiparticle description is absent. In such situations, however, the many-body Kubo formula from linear response theory, involving matrix elements of the current operator computed with many-body wavefunctions, remains valid. Working directly in the many-body Hilbert space and not making any reference to quasiparticles (or lack thereof), we address the puzzle of linear in temperature ($T$-linear) resistivity seen in non-Fermi liquid phases that occur in several strongly correlated condensed matter systems. We derive a simple criterion for the occurrence of $T$-linear resistivity based on an analysis of the contributions to the many-body Kubo formula, determined by an energy invariant ``$f$-function" involving current matrix elements and energy eigenvalues that describes the DC conductivity of the system in the microcanonical ensemble. Using full diagonalization, we test this criterion for the $f$-function in the spinless nearest neighbor Hubbard model and in a system of Sachdev-Ye-Kitaev dots coupled by weak single particle hopping. We also study the $f$-function for the spin conductivity in the 2D Heisenberg model and arrive at similar conclusions. Our work suggests that a general principle, formulated in terms of many-body Hilbert space concepts, is at the core of the occurrence of $T$-linear resistivity in a wide range of systems, and precisely translates $T$-linear resistivity into a notion of energy scale invariance far beyond what is typically associated with quantum critical points.
\end{abstract}
\maketitle

{\it{Introduction:}}
How do strongly correlated materials (eg. the high $T_c$ superconducting cuprates, heavy fermions and more recently, 
twisted bilayer graphene~\cite{Bednorz1986,Cooper603,TailleferSM,Legros18,Keimer2015,Stewart_review,Bruin804,Cao_Planckian}) conduct electricity at finite temperature? This is a fundamental question that has existed since the realization of these materials, and the inception of this field decades ago. Experiments have helped build an intricate picture of the phases that occur, both from the point of their static and dynamical properties at finite temperature, but much remains to be accomplished in order to have a definitive theoretical understanding of these materials. For example, at and close to 
optimal doping, the superconducting phase transitions to the ``non Fermi liquid" (nFL) or ``strange metal" phase which is characterized by an electrical resistivity that scales linearly with temperature ($T$-linear) over a wide range of $T$~\cite{Cooper603,TailleferSM,Legros18,Keimer2015,Bruin804,Stewart_review, Giraldo_Gallo_Science}. This is in sharp contrast to Fermi liquid (FL) theory which predicts that the electrical resistivity of a metal scales as $T^2$ \cite{Ziman}. 

nFLs, in contrast to FLs, are characterized by a lack of quasiparticles, leading to a concerted effort to find models and mechanisms by which $T$-linear resistivity can occur. Prominent among these is the Sachdev-Ye-Kitaev (SYK) model and its variants~\cite{Sachdev_Ye,SachdevComplexSYK,Kitaev2015talk} which are analytically solvable in a large $N$ limit and exhibit $T$-linear resistivity (when multiple SYK dots are coupled)~\cite{Song_SYK,ParcolletGeorges}. However, the connection of this model to a realistic microscopic model remains to be established. Recent experiments with cold atoms~\cite{Bakr_2019} have shown the existence of $T$-linear resistivity in the Hubbard model which has been supported by dynamical mean field theory \cite{Vranic2020,Vlad2015,Deng2013,Georges_DMFT,PK1998} and exact diagonalization \cite{Kokalj2017,Vranic2020} calculations. 

\begin{figure*}
\subfigimg[height=135pt]{
\hspace{-6pt}\textsf{\scriptsize(a)}
}{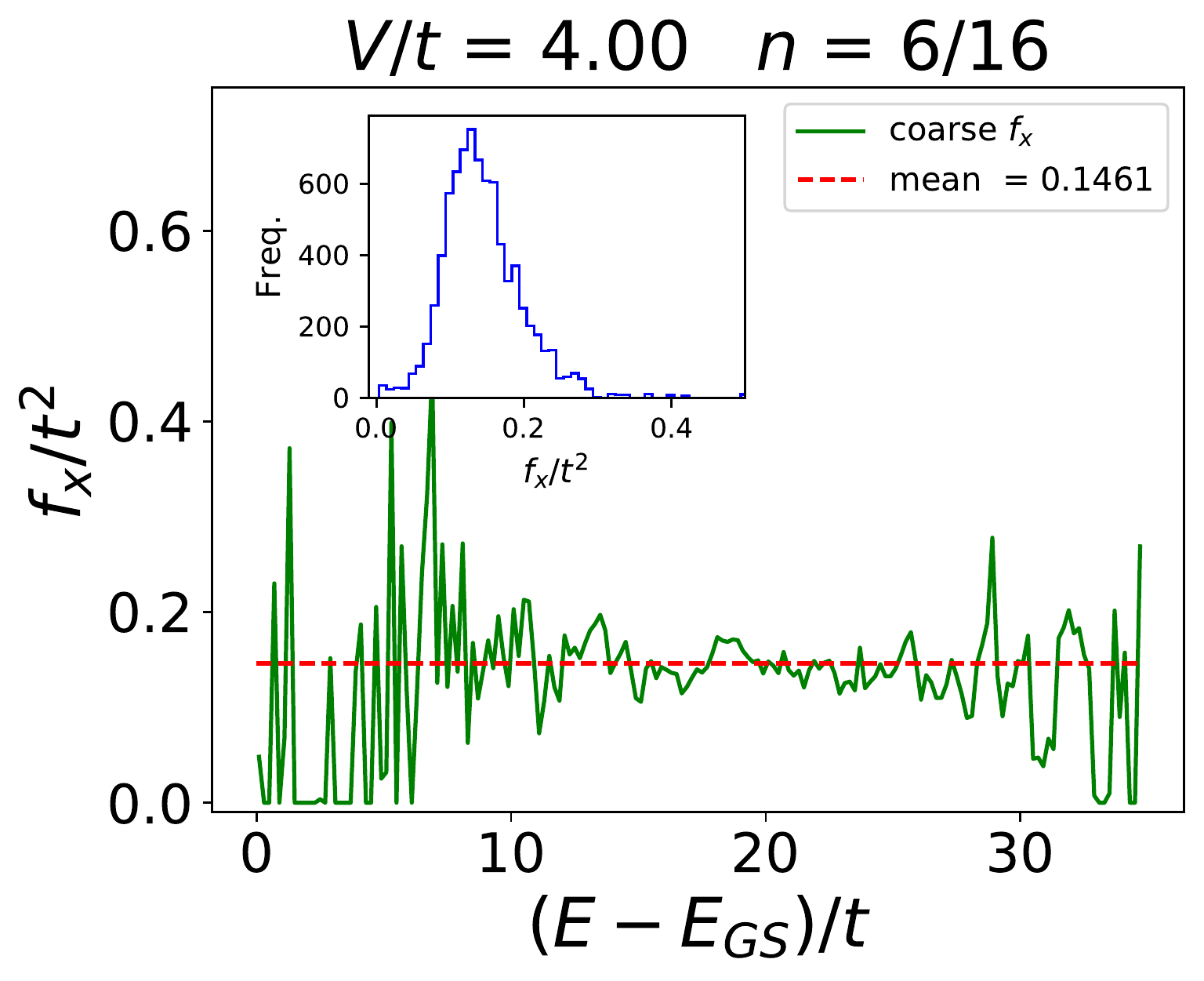} \subfigimg[height=135pt]{
\hspace{-6pt}\textsf{\scriptsize(b)}
}{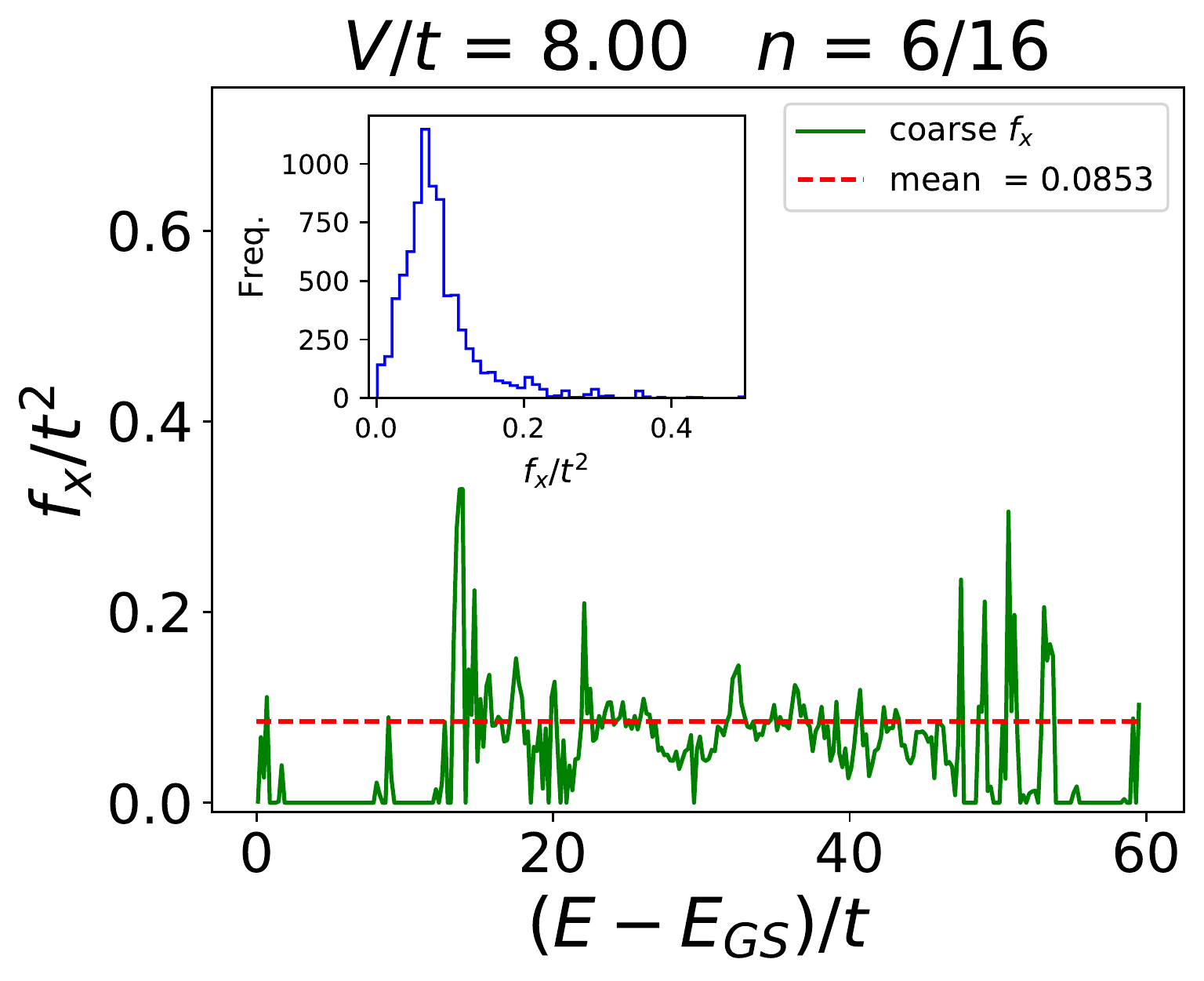}
\subfigimg[height=135pt]{
\hspace{-6pt}\textsf{\scriptsize(c)}
}{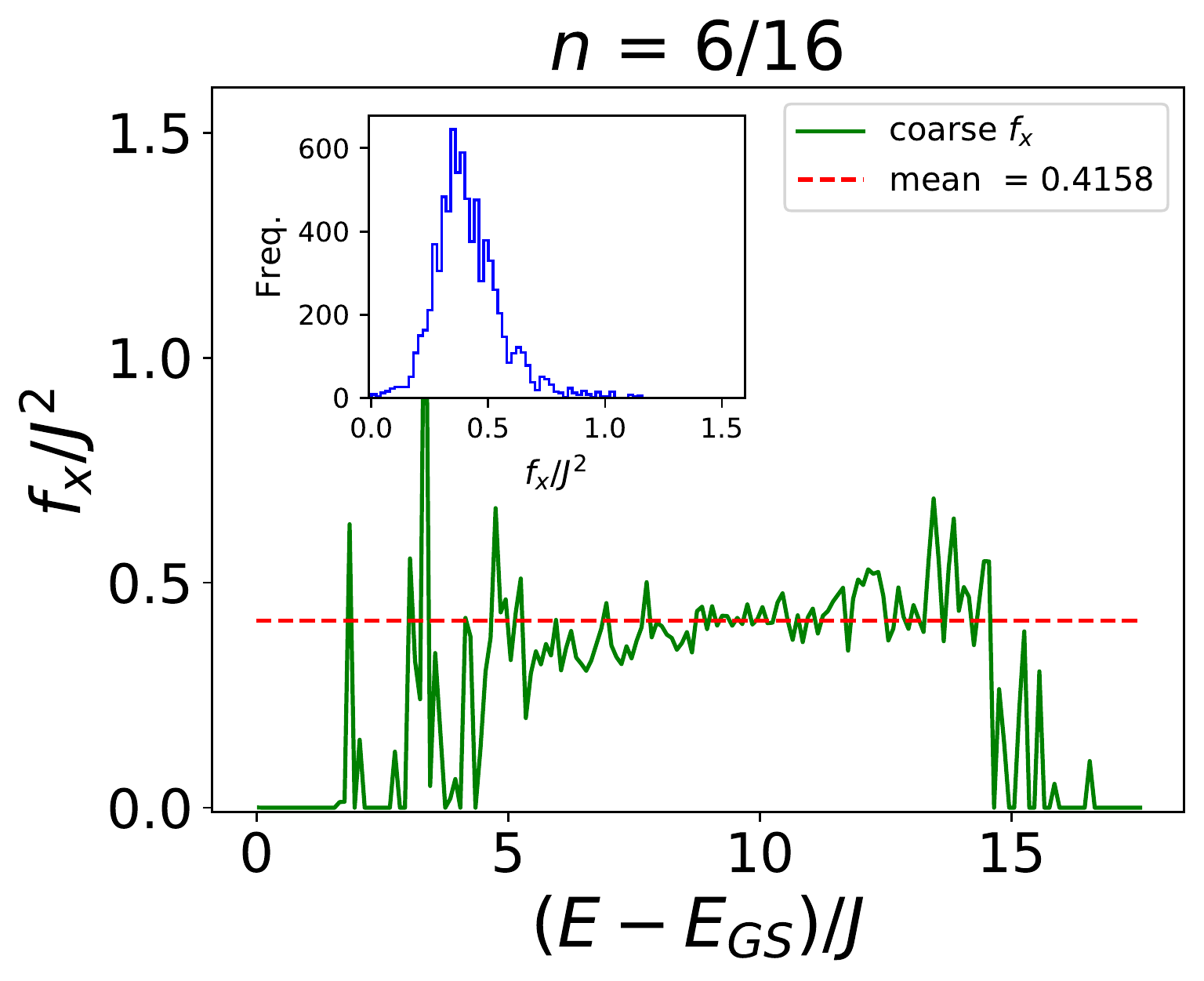}
\caption{(a), (b) $f_x(E)$ for the charge conductivity, using $\eta=0.20 t$, as a function of the energy $E$ (in units of $t$) for the $4\times4$ 2D square lattice nearest neighbor spinless Hubbard model for $V/t = 4,8$ and for a filling of $n=6/16$. In each case, the ground state energy $E_{GS}$ has been subtracted out. The insets show histograms of $f_x(E)$ values with the bin width set to 0.01. (c) $f_x(E)$ for the spin conductivity,using $\eta=0.10 J$, 
of the 2D spin $1/2$ nearest-neighbor Heisenberg model (mapped to a hardcore-bosonic model) on a $4\times4$ square lattice with (bosonic) filling $n=6/16$. }
\label{fig:ffn_spinless_Hubbard}
\end{figure*}

We address the question of $T$-linear resistivity, circumventing the issue of quasiparticles (or lack thereof) 
completely. We work directly with the full set of quantum many-body wavefunctions (which contain information about the resistivity at all temperature scales), and appeal to a direct analysis of the many-body Kubo formula~\cite{Kubo_1957}. This is valid within linear response, which is sufficient given that the experimentally applied electric fields are small perturbations to the full electronic Hamiltonian. The expression for the longitudinal conductivity ({\it i.e.} the inverse of the electrical resistivity $\rho_{\alpha}$) is given by,
\begin{equation}
	\sigma_{\alpha} (\omega,T) = \pi \frac{1-e^{-\beta \omega}}{\omega Z} \sum_{n,m} \frac{|I^{\alpha}_{nm}|^2}{e^{\beta E_n}} \delta(E_n + \omega - E_m ) 
\label{eq:kubo}
\end{equation}
where $\omega$ is the energy of interest (the DC limit corresponds to $\omega \rightarrow 0$), $E_n$, $E_m$ are eigenenergies of the $n^{th}$ and $m^{th}$ eigenstates respectively, $Z$ is the partition function, $\alpha$ is a label for the spatial direction ($x$ or $y$ in two dimensions) and $I^{\alpha}_{nm} \equiv \langle n | I^{\alpha} | m \rangle $ are matrix elements of the current operator, and $\beta$ is the inverse temperature.

At extremely high $T$ (higher than the many-body bandwidth), Ref.~\cite{Mukerjee_2006} stated a straightforward 
reason for $T$-linear resistivity. In this limit, the thermal factors $\exp(-\beta E_n)$, occurring in the numerator and the partition function in the denominator, all become one. At high temperature and vanishing frequency, $\beta \rightarrow 0$ and $\omega \rightarrow 0$, the factor of $(1-\exp(-\beta \omega))/\omega \rightarrow \beta$, which yields the linear in  $\beta$ conductivity and hence $T$-linear resistivity. Though mathematically appealing, this argument alone does not explain why $T$-linear resistivity remarkably survives to lower $T$. Studies of the high $T$ limit by Refs.~\cite{Perepelitsky, Lindner_Auerbach, Gunnarsson2003} also suggested that many aspects of $T$-linear behavior can be understood from high temperature expansions.

Our key contribution is to establish a criterion for $T$-linear resistivity at finite temperature and to test its general validity. We note that the Kubo formula can be rewritten as, 
\begin{equation}
\sigma_{\alpha} (\omega,T) = \Big( \frac{1-e^{-\beta \omega}}{\omega} \Big) \Big( \frac{\sum_{n} e^{-\beta E_n} f_{\alpha}(E_n, |n\rangle, \omega)} {\sum_{n} e^{-\beta E_n}} \Big),
\label{eq:kubo2}
\end{equation}
where we have introduced the $f$-function, defined as,
\begin{subequations}
\begin{eqnarray}
	f_{\alpha}(E_n, |n\rangle,\omega) &\equiv& \pi \sum_{m} |I^{\alpha}_{nm}|^2 \delta(E_n + \omega - E_m ) \\
	&\equiv& \textrm{lim}_{\eta \rightarrow 0 } \sum_{m} \frac{\eta |I^{\alpha}_{nm}|^2}{\eta^2 + (E_n + \omega - E_m)^2},~ \label{eq:kubo3}
\end{eqnarray}
\end{subequations}
where $\eta$ is a broadening parameter whose use is necessitated by the discreteness of the many-body spectrum in numerical computations on a finite sized system. Once again for $\omega \rightarrow 0$, the prefactor outside the summation yields the desired factor of $\beta$. This means that the remaining terms must conspire to \textit{perfectly} cancel out to have no temperature dependence. This can happen for an arbitrary range of $T$, if $f(E_n, |n\rangle) \equiv f(E_n, |n\rangle, \omega \rightarrow 0)$ is constant, {\it i.e}. independent of the energy of the eigenstate and the eigenstate itself. Since there is a continuum of many-body energies and eigenstates in the thermodynamic limit, it is meaningful to coarse grain the $f$-function by simple averaging within a narrow energy window, as long as the energy window over which the averaging is done is significantly smaller than the lowest temperature scale of interest: 
\begin{equation}
f_\alpha(E) \equiv \frac{1}{g(E)}\sum_n \delta(E_n-E) f_{\alpha}(E_n, |n\rangle),   
\label{eq:coarse_f}
\end{equation}
where $g(E)\equiv\sum_{n}\delta(E_n-E)$ is the many-body density of states ($f_\alpha(E_n,|n\rangle) = f_\alpha(E_n)$ follows from the eigenstate thermalization hypothesis (ETH)~\cite{SM}, but the coarse-grained function is well-defined even in situations where ETH does not hold). 

For this averaged $f_{\alpha}(E)$, we show \cite{SM} that its energy invariance is the only generic possibility for $T$-linear resistivity at arbitrary temperature. This condition must hold in situations where the slope $d\rho_\alpha/dT$ has been found to be invariant with temperature \cite{Cha_2020}. Furthermore, for resistivity scaling as other powers of $T$, there does not appear to be any such generic invariant that is defined in the microcanonical ensemble, which indicates that exact $T$-linear resistivity is somehow ``special". 

The $f$-function recasts the complex finite temperature problem into an analysis of the quantum mechanical energies and matrix elements of the current operator. In realistic models, we may expect only \textit{approximate} $T$-linear resistivity, in which case the conditions on the $f$-function can be somewhat relaxed: we then expect $f(E)$ to be constant only in energy regimes corresponding to temperatures where the $T$-linear contribution to the resistivity dominates. At low energies we may expect to see physics associated with antiferromagnetism, superconductivity or FL behavior, and the $f$-function can not be constant in these regimes. 

To test our assertions, we carry out a systematic numerical investigation of the $f$-function in the spinless Hubbard and SYK models. We also pose and answer an analogous question about spin conductivity in the two dimensional spin-1/2 square lattice Heisenberg model. 

{\it{Spinless nearest neighbor Hubbard model:}}
Consider a nearest neighbor (nn) spinless Hubbard model on the 2D square lattice, 
\begin{equation}
	H = -t \sum_{\langle i,j \rangle} c^{\dagger}_i c_j  + \textrm{h.c.} + V \sum_{\langle i,j \rangle} n_i n_j,
\end{equation}
where $\langle i,j \rangle$ refer to nn pairs, $t$ is the nn hopping (which we set to 1 for our calculations), $V$ is the strength of the nn repulsion and $c^{\dagger}_i$ and $c_i $ are the usual electron creation and destruction operators. $n_i = c^{\dagger}_i c_i$ is the number operator. The current operator is defined as,
\begin{equation}
	I^{x (y)} =  \frac{it}{\sqrt{N_s}} \sum_{j=1}^{N_s} (c_{j+ \hat{x} (\hat{y})}^{\dagger}c_j - c_{j}^{\dagger}c_{j+ \hat{x} (\hat{y})}), 
\end{equation}
where $N_s$ is the total number of sites. We simulate an isotropic lattice ($4 \times 4$ torus i.e. periodic boundary conditions in both directions), and plot only $f_x$ (computed from $I^x$), since $f_y$ (computed from $I^y$) is identical.  

\begin{figure}[ht]
\includegraphics[width=0.48\textwidth]{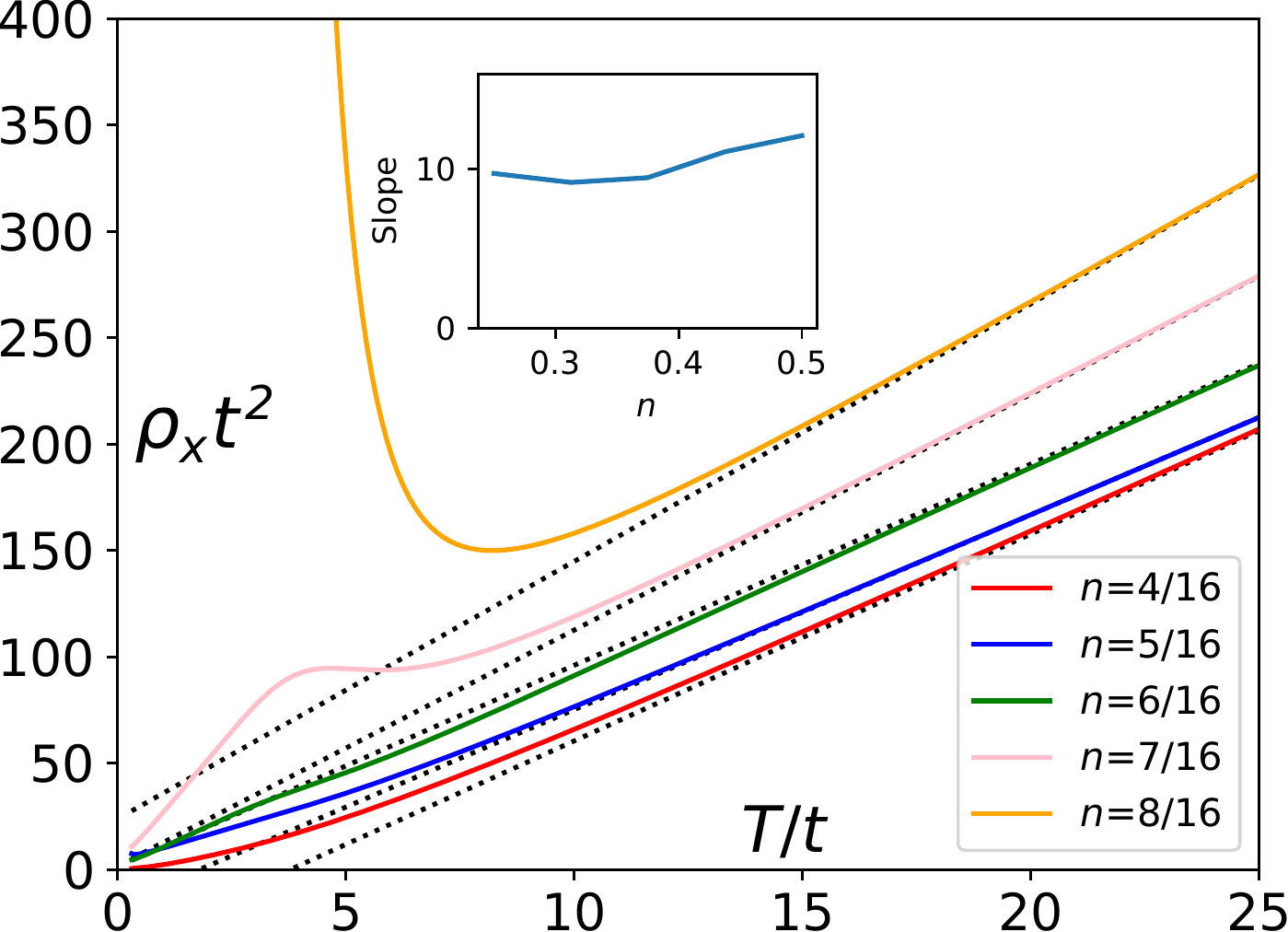}
\caption{Resistivity ($\rho_x t^2$), using $\eta = 0.10 t$, for the spinless Hubbard model for a representative value of interaction $V/t=8$ and various fillings. The high temperature part of the curves ($T=30t$ to $70t$, not shown) is fit to a linear function, the corresponding slope is shown in the inset. The low temperature physics is characterized by metallic or insulating phases which show clear deviations from $T$-linear resistivity.}
\label{fig:rho_spinless_Hubbard}
\end{figure}

Fig.~\ref{fig:ffn_spinless_Hubbard} (a) and (b) show plots of $f_x (E)$, for representative values of $V/t$ at a filling of $n=6/16$. The ground state energy in each case has been subtracted out on the energy axis. A broadening parameter of $\eta = 0.2 t$ is used.%, results of other broadening parameters ($\eta = 0.1t,0.3t$) for similar fillings are shown in the Supplementary Material (SM) \cite{SM}. 
Additionally, the energy axis is split up into bins of size $\eta$ and the coarsened value of $f_x (E)$ is obtained by simple averaging over all the eigenstates with eigenenergies that lie in a given bin, as in Eq.~\eqref{eq:coarse_f}. The mean value of $f_x (E)$ averaged over the entire eigenspectrum is also shown as a guide to the eye. 

If one focuses on the center of the many-body spectrum, $f_x (E)$ does appear to be remarkably flat for all the cases shown. To quantify the degree of flatness of the $f$-function, we plot the histogram of $f_x(E_k,|k\rangle)$ values for all eigenstates $|k\rangle$ in the spectrum (assigning degenerate states the same $f_x$-value) in the inset.
We observe that the $f$-value is indeed peaked around a typical value. (In the Supplementary Material (SM)~\cite{SM} we also show the $f$-function for other fillings, interaction strengths and broadening parameters.)

Fig.~\ref{fig:rho_spinless_Hubbard} shows a representative set of resistivity curves for $V/t=8$ and different particle fillings. For small fillings and low temperature, one has a dilute gas of well defined electronic quasiparticles, the $f$-function is high at low energies, correspondingly the resistivity shows deviation from $T$-linear behavior that is present at large $T$. At half filling, one has insulating behavior at low temperature expected of the charge density wave phase. The slope of the $T$-linear portion (obtained by biasing the fit to include only high $T$) is shown in the inset and is approximately (but not exactly) constant with filling.

\begin{figure}[ht]
\includegraphics[width=0.48\textwidth]{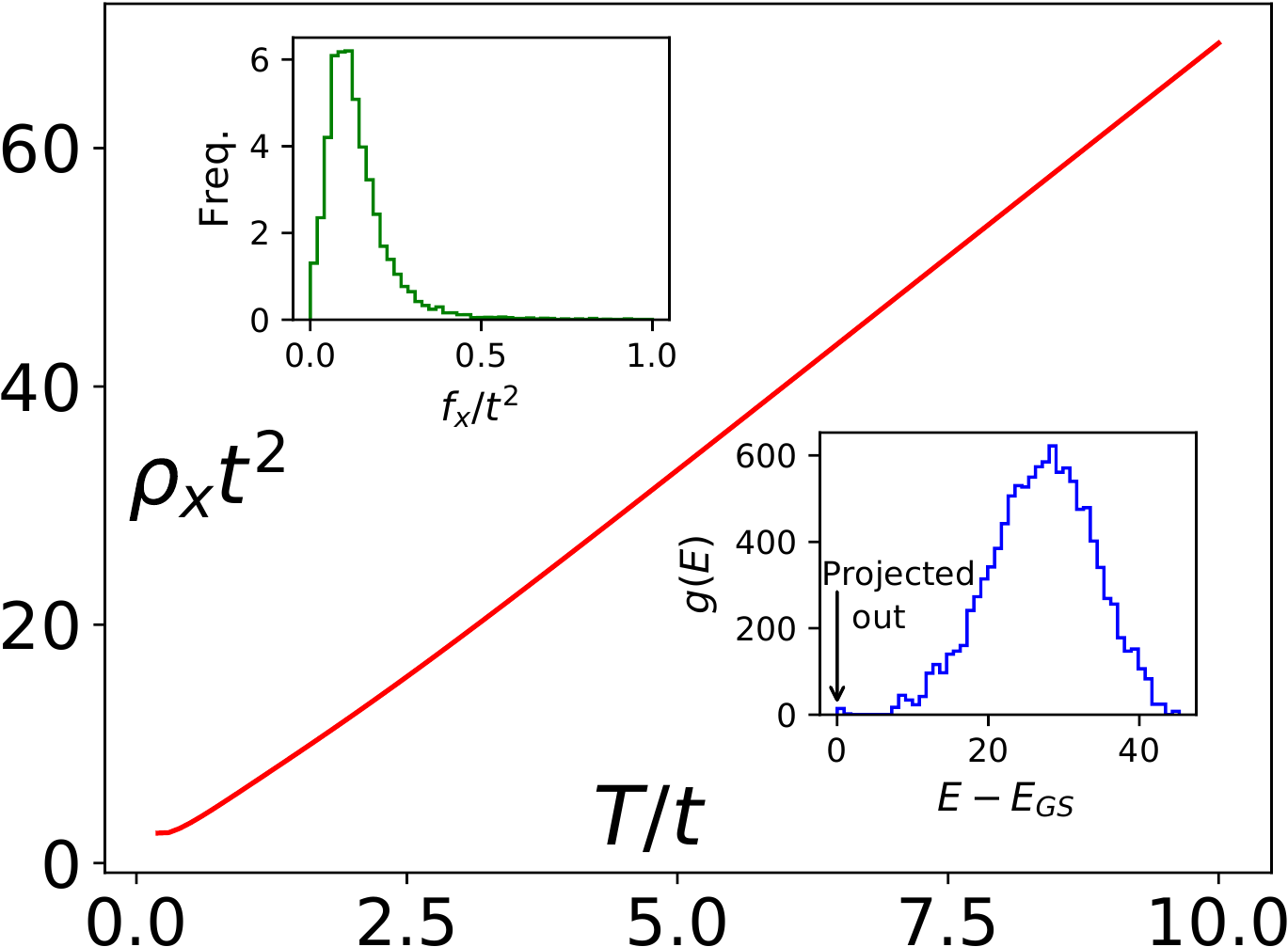}
\caption{Resistivity ($\rho_x t^2$), using $\eta = 0.10 t$, for the spinless Hubbard model for $V/t=4.3$ and filling 7/16, after the lowest energy manifold of 16 states is projected out of the calculation. The inset at the bottom right shows the density of states $g(E)$. The inset at the top left shows the histogram of $f_x$ values (normalized histogram) for the remaining $11440-16 = 11424$ eigenstates.}
\label{fig:projected_rho_spinless_Hubbard}
\end{figure}

We now demonstrate that the operative mechanism behind $f$-invariance in the spinless nearest neighbor Hubbard model stems from an incoherent quantum liquid of states that extends across energy scales. To do this, we consider the model near half-filling, where the incoherent quantum liquid is separated from the low energy manifold of states $|L_n\rangle$ by a gap in the many-body spectrum. We proceed to project out this low energy manifold by redefining $H\rightarrow H + \infty\sum_n|L_n\rangle\langle L_n|$ \footnote{While this introduces a corresponding infinite term in the current operator, such a term does not contribute to transport, as it has matrix elements only between infinite energy states, whose weight in the Kubo formula for the conductivity is a polynomial times an exponential, {\it i.e.} $\infty^m e^{-\beta\infty}$, which vanishes.} (Fig.~\ref{fig:projected_rho_spinless_Hubbard}). Doing so makes the incoherent quantum liquid extend all the way down to low energies \footnote{The idea of the extension of incoherent behavior down to low energies by introducing strong long range interactions was also used in Ref. \cite{PatelPlanckian}}. Then, in the strongly correlated regime $V/t \gtrsim 1$, we find that $T$-linear resistivity extends from high $T$ down to nearly $T=0$ without a slope change (Fig. \ref{fig:projected_rho_spinless_Hubbard}), and the resistivity at low $T$ is not much larger than $1/t^2$, {\it i.e.} not bad metallic. Consequently, $f$-invariance now extends across the energy spectrum in the modified model. 

This projection procedure also causes the transfer of single-particle spectral weight from the upper Hubbard bands down to low frequencies (for a detailed discussion, see SM). The resulting UV-IR mixing in the local single-particle spectral function \cite{Mottness, MottnessRMP} therefore results in energy scale invariant behavior with respect to the addition or removal of a single particle. Our calculations suggest that an analogous situation arises with the current operator; under projection it also redistributes the UV spectral weight down to low energies, which in turn extends the $T$-linear regime to low $T$.

{\it{Heisenberg model:}}
The fermionic nature of the constituents has no particular relevance in our Hilbert space viewpoint. This motivates an investigation of a very different model, the 2D spin-$1/2$ Heisenberg model on a square lattice with the Hamiltonian,
\begin{equation}
	H = J \sum_{\langle i,j \rangle} \bf{S}_{i} \cdot \bf{S}_{j},
\end{equation}
from the point of view of its spin conductivity. ($\bf{S_i}$ represent the usual spin-$1/2$ operators on site $i$). The spin current is defined as $I^{x (y)}_S = iJ\sum_{j=1}^{N_s} (S_{j+x (y)}^{+}S_j^{-} - S_{j}^{+}S_{j+x (y)}^{-})/\sqrt{N_s}$. \cite{Sentef} (We set $J=1$ in our calculations.) 
The model maps to one of hardcore bosons with $t=-J/2$ and $V=J$; a previous investigation by Ref.~\cite{Lindner_Auerbach} using high temperature expansions showed that such particles also show $T$-linear resistivity. 

We evaluate the $f$ function for the $4 \times 4$ torus in different magnetization sectors, equivalent to different fillings of hardcore bosons. We find that the $f$-function is indeed flat when viewed at intermediate energy (see Fig.~\ref{fig:ffn_spinless_Hubbard} (c) for a representative calculation), paralleling our observations for the fermionic case. These findings hint at the diminished role of particle statistics at high temperature, which we find remarkable yet consistent with recent experiments that have suggested the occurrence of a ``bosonic strange metal"~\cite{Yang2022} with robust $T$-linear resistivity. It remains to be seen if this effect can be observed for the ``spin-resistivity" in insulating magnetic materials.

{\it{SYK model:}} 
Finally, we discuss our results for the zero-dimensional SYK model of spinless complex fermions $c_i$ \cite{SachdevComplexSYK,Kitaev2015talk}. 
Its Hamiltonian is, 
\begin{equation}
	H = \frac{1}{(2N)^{3/2}}\sum_{i,j,k,l=1}^N J_{ijkl} c^{\dagger}_i c^{\dagger}_j c_k c_l,
\end{equation}
where $J_{ijkl}$ are independent random complex numbers chosen from a Gaussian distribution with standard deviation $J$, and the model is defined in the limit of large $N$.

\begin{figure}[ht]
\includegraphics[width=0.48\textwidth]{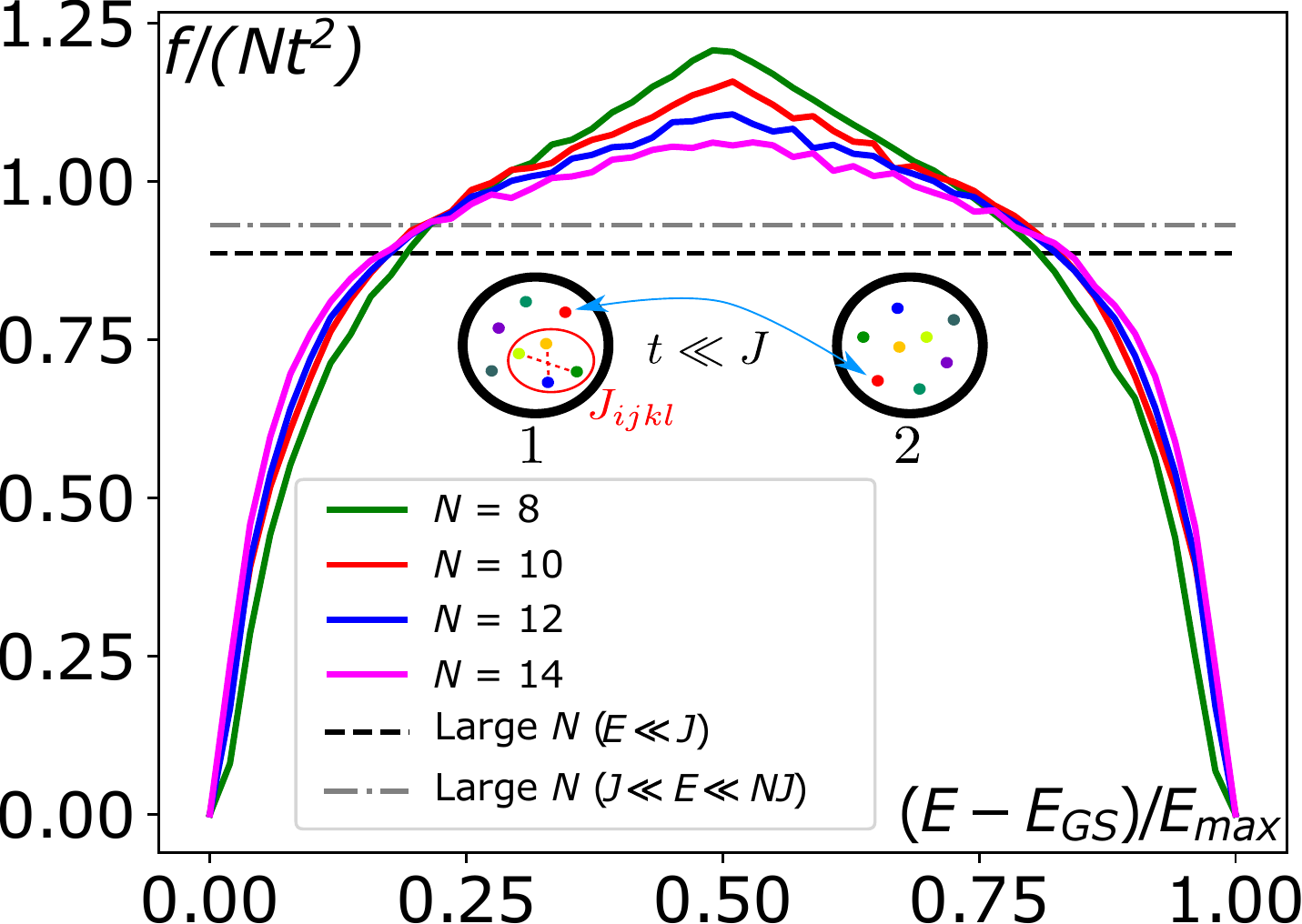}
\caption{$f$-function computed for a two SYK dot system (each having $N$ sites) using exact diagonalization for various values of $N$, averaged over 128 realizations each. The many-body bandwidth is $E_{\mathrm{max}}$, and the ground state energy $E_{\mathrm{GS}}$ has again been subtracted out (for each disorder realization). The dashed lines indicate $f$ derived from the large $N$ results for $\rho(T)$ obtained in previous work \cite{Cha_2020}, where we take $f\approx \rho(T)/T$ \cite{SM}. The curves will end up within the region between the dashed lines as $N$ is made large.}
\label{fig:syk_fig}
\end{figure}

Owing to the high amount of frustration, the model inhibits the formation of ordered states \cite{Sachdev_Ye} in the limit of $N \rightarrow \infty$ Moreover, the fully random interactions and the absence of single particle hopping also means that there is no Fermi liquid or glassy phase down to zero energy (zero temperature) \cite{GPS2001,SachdevComplexSYK,GurAri2018,BaldwinSwingle}. Thus, the SYK model is one of the simplest known models where nFL physics persists all the way down to $T=0$. 

The concept of charge transport is not well defined for a single zero-dimensional SYK dot. However, one can weakly couple SYK dots (labeled $1,2$) with infinitesimal single-particle hopping $t$ (Fig. \ref{fig:syk_fig}), and define an appropriate current operator $I = it\sum_{j=1}^N(c^\dagger_{j,1}c_{j,2}-c^\dagger_{j,2}c_{j,1})$, we drop the direction label $\alpha$ here. Field theoretic calculations in the large $N$ limit, where $T\ll NJ$ by definition, have revealed that the resistivity $\rho$ is linear in $T$ \cite{ParcolletGeorges,Song_SYK}, and its slope $d\rho/dT$ is nearly invariant \cite{Cha_2020}, {\it i.e}. it does not depend on the temperature scale (with respect to $J$) that the system is at, even though the temperature dependences of other physical quantities change drastically as $T$ is increased past $J$ \cite{Cha_2020} (such as the compressibility, which changes from $\sim T^0$ to $\sim T^{-1}$).

We compute the $f$-function of the two dot system as follows: since the hopping $t$ is infinitesimal, the dots are effectively decoupled, and the many body states $|n\rangle=|n_1\rangle|n_2\rangle$ are therefore (fermionic) product states of the states on the individual dots. We exactly diagonalize the Hamiltonians for the two dots individually, which have uncorrelated realizations of $J_{ijkl}$. We then have ($E_m,Q_m\equiv E_{m_1},Q_{m_1}+E_{m_2},Q_{m_2}$)
\begin{align}
&f(E) = \frac{2\pi t^2}{g(E)}\sum_{n_1,n_2}\sum_{m_1,m_2}\delta(E_m-E)\delta(E_n-E)\delta_{Q_m,Q} \nonumber \\
&\times\delta_{Q_n,Q}\Bigg|\sum_{i=1}^N\langle n_1|c_{i,1}|m_1\rangle\langle n_2|c^\dagger_{i,2}|m_2\rangle(-1)^{Q_{m_1}}\Bigg|^2,
\end{align}
where $g(E)$ is the many-body density of states of the two dot system, and the total charge on the two dots is $Q$.

Fig.~\ref{fig:syk_fig} shows the results of our calculations at $Q=N$ for $N=8$ to $N=14$, which were obtained after averaging over 128 realizations of $J_{ijkl}$ for each $N$. We find that in the middle of the spectrum, the $f$-function tends to get flatter with increasing $N$, approaching the large $N$ result. Towards the edges, the $f$-values are smaller, but increase towards the large $N$ result with increasing $N$: thus, the profile of the $f$-function appears to be asymptoting towards the nearly invariant large $N$ result as $N$ is increased. Remarkably, the $f$-invariance also appears to extend to energy scales $E\sim NJ$ in the middle of the band, far higher than those accessed in the large $N$ field theory calculations, where $E\ll NJ$ by definition. 

{\it{Discussion:}} 
We conclude by discussing the implications of the energy invariance of the $f$-function, which is a purely microcanonical quantity. For this energy invariance to occur, a subtle interplay between the average size of the matrix element of the current operator and the available number of many-body states at a given energy density is required. The energy invariance of the $f$-function encodes a notion of energy scale invariance across the many-body spectrum, which is far beyond the purview of low energy effective field theories. Importantly, when viewed in terms of the many-body Hilbert space, different models suggest a universal mechanism behind $T$-linear resistivity.

Certain correlated electron materials display ``perfect" $T$-linear resistivity across multiple decades of temperature \cite{Anderson1992,GurvitchFiory1987,MartinFiory1990,TakagiBatlogg1992}, which is often associated with the presence of a quantum critical point \cite{SK2011,Sachdev_QPT}. This resistivity goes from $\rho\ll h/e^2$ at low $T$, to a ``bad metal" regime where $\rho \gg h/e^2$ at high $T$, in which the classical mean free path of the electrons becomes comparable to a lattice spacing \cite{EmeryKivelson}. This suggests very different physics in the two regimes \cite{Bakr_2019,Patel2018}, yet $f$-invariance must hold across the crossover between them, indicating that they are still related. To probe this physics further, larger system sizes are required: it would be interesting to construct variational wavefunctions \cite{Ferrari2018} or matrix product states for excited states in models of quantum critical metals \cite{BergMC} that could capture this crossover, and see how $f$-invariance can manifest in terms of the physical parameters used to define these wavefunctions. Also, other computational strategies based on shift-invert based algorithms that target states at a given energy density could be used for calculation of the $f$-function for larger system sizes, and thus shed further light on the problem of $T$-linear resistivity. 

{\it{Acknowledgements:}} 
We thank V. Dobrosavljevic, M. Randeria, S. Sachdev, and Y. Wan for discussions on transport and bad metals, and S. Hartnoll, J. McGreevy, S. Sachdev, and S. Shastry for useful comments on the paper. H.J.C. thanks O. Vafek for pointing him to Ref.~\cite{Mukerjee_2006}, which stimulated his interest in the subject. A.A.P. was supported by the Miller Institute for Basic Research in Science. H.J.C. was supported by NSF CAREER Grant No. DMR 2046570 and startup funds from Florida State University and the National High Magnetic Field Laboratory. The National High Magnetic Field Laboratory is supported by the National Science Foundation through NSF/DMR-1644779 and the state of Florida. We thank the Research Computing Cluster (RCC) and the Planck cluster at Florida State University for computing resources.

A.A.P and H.J.C. contributed equally to this work. 

\bibliography{refs}

\clearpage
\newpage
%\beginsupplement
\setcounter{equation}{0}
\setcounter{figure}{0}
\setcounter{table}{0}
\setcounter{page}{1}
\makeatletter
\renewcommand{\theequation}{S\arabic{equation}}
\renewcommand{\thefigure}{S\arabic{figure}}
\renewcommand{\bibnumfmt}[1]{[#1]}
\renewcommand{\citenumfont}[1]{#1}

%%%%%%%%%% Merge with supplemental materials %%%%%%%%%%
\widetext
\begin{center}
\Large{Supplementary Material for ``Many-body energy invariant for $T$-linear resistivity"}
\end{center}
%\author{Aavishkar A. Patel}
%\email{aavishkarpatel@berkeley.edu}
%\affiliation{Department of Physics, University of California, Berkeley, CA 94720, USA}
%\author{Hitesh J. Changlani}
%\email{hchanglani@fsu.edu}
%%\thanks{Both authors contributed equally to this work.}
%\affiliation{Department of Physics, Florida State University, Tallahassee, Florida 32306, USA}
%\affiliation{National High Magnetic Field Laboratory, Tallahassee, Florida 32310, USA} 
%\maketitle
%
\section{Proof of $f$-invariance given $T$-linear resistivity at all temperatures}

In the manuscript we wrote the Kubo formula for the dynamical conductivity (along direction $\alpha$) as, 
\begin{equation}
\sigma_{\alpha} (\omega,T) = \Big( \frac{1-e^{-\beta \omega}}{\omega} \Big) \Big( \frac{\sum_{n} e^{-\beta E_n} f_{\alpha}(E_n, |n\rangle, \omega)} {\sum_{n} e^{-\beta E_n}} \Big)
\end{equation}
where the $f$-function was defined as,
\begin{subequations}
\begin{eqnarray}
	f_{\alpha}(E_n, |n\rangle,\omega) &\equiv& \pi \sum_{m} |I^{\alpha}_{nm}|^2 \delta(E_n + \omega - E_m ) \\
	&\equiv& \textrm{lim}_{\eta \rightarrow 0 } \sum_{m} \frac{\eta |I^{\alpha}_{nm}|^2}{\eta^2 + (E_n + \omega - E_m)^2}.
\end{eqnarray}
\end{subequations}

For most generic quantum models, the eigenstate thermalization hypothesis (ETH)~\cite{Srednicki1994, Deutsch1991} is expected to hold in the thermodynamic limit. Since the current operator involves only local terms, we expect that the $f$-function should not depend too sensitively on which state $|n\rangle$ is used, as long as it is at a given energy $E$. Under these conditions, the $f$-function is a function of energy alone. Even in situations where ETH does not hold, it is meaningful to coarse grain the $f$-function by simple averaging, as long as the energy window over which the averaging is done is significantly smaller than the temperature scale of interest. We therefore study $f_{\alpha}(E_n, |n\rangle) \equiv f_{\alpha}(E_n, |n\rangle, \omega=0)$ and coarse grain it by defining,
\begin{equation}
f_\alpha(E) \equiv \frac{1}{g(E)}\sum_{n}\delta(E_n-E)f_{\alpha}(E_n, |n\rangle),   
\end{equation}
where $g(E)\equiv\sum_n\delta(E_n-E)$ is the many-body density of states.

Working with this coarse-grained function and dropping the subscript $\alpha$, we now show that exact $T$-linear DC resistivity for \textit{all} $T$ implies that $f(E)$ is independent of energy. First, we can easily see that if $g(E)$ has support only at one value of $E$, then 
\begin{equation}
T\sigma(0,T) = \frac{\sum_n g(E_n) f(E_n) e^{-\beta E_n}}{\sum_n g(E_n)e^{-\beta E_n}},
\end{equation}
is automatically independent of $T$, leading to exact $T$-linear DC resistivity. If $g(E)$ has support at two values of $E$, $E_1$ and $E_2$, then the condition $d(T\sigma(0,T))/d\beta = 0$ gives $f(E_2) = f(E_1)$, and thus exact $T$-linear resistivity. We can now generalize this to a spectrum that has support at an arbitrary number of values $N_E > 2$ of $E$. The conditions $d(T\sigma(0,T))/d\beta = 0,~...~,d^{N_E-1}(T\sigma(0,T))/d\beta^{N_E-1}=0$ yield a linear system of equations for $f(E_2),~...~,f(E_{N_E})$, which may be represented by a matrix $M$ with entries $M_{mn} = d^mW_n/d\beta^m$, where $W_n = g(E_{n+1})e^{-\beta E_{n+1}}/(\sum_k g(E_k)e^{-\beta E_k})$, as $\sum_{n=1}^{N_E-1} M_{mn}f(E_{n+1}) = -(d^mW_0/d\beta^m)f(E_1)$. 

Then, the columns of $M$ are linearly independent, as linear dependence would require that $\Xi = \sum_{n=1}^{N_E-1} a_n W_n$, {\it i.e.} $\Xi = \left(\sum_{n=1}^{N_E-1} a_n g(E_{n+1})e^{-\beta E_{n+1}}\right)/(\sum_k g(E_k)e^{-\beta E_k})$, be independent of $\beta$, as $d\Xi/d\beta = 0$. This is not possible, because while the sum in the denominator contains a term proportional to $e^{-\beta E_1}$, {\it i.e.} $g(E_1)e^{-\beta E_1}$, there is no term proportional to $e^{-\beta E_1}$ present in the sum in the numerator, and it is not possible to express an exponential with a given decay rate as a linear combination of exponentials with decay rates different from the given rate. Therefore, the matrix $M$ is invertible, and the linear system is thus solved with a unique solution of $f(E_2)=...=f(E_{N_E}) = f(E_1)$.

A few words of caution are now in order. For \textit{approximate} $T$-linear resistivity, the conditions on the $f$-function can be relaxed, it does not strictly have to be energy invariant anymore. For example, there could be ``scars" in the spectrum (one or many atypical states which do not satisfy ETH)~\cite{Serbyn_review,Lee_PRBR2020} in an otherwise chaotic spectrum. As long as their contribution to the partition function and hence the conductivity, is measure zero in the thermodynamic limit, our general statement about the $f$-function should hold. Additionally, realistic models are not expected to have a perfectly constant $f$-value, especially at the edges of their many-body spectra. At low energy (temperature) we expect to see physics associated with antiferromagnetism, superconductivity or FL, thus the $f$-function can not be constant in this regime. 

\begin{figure}
\includegraphics[width=0.33\linewidth]{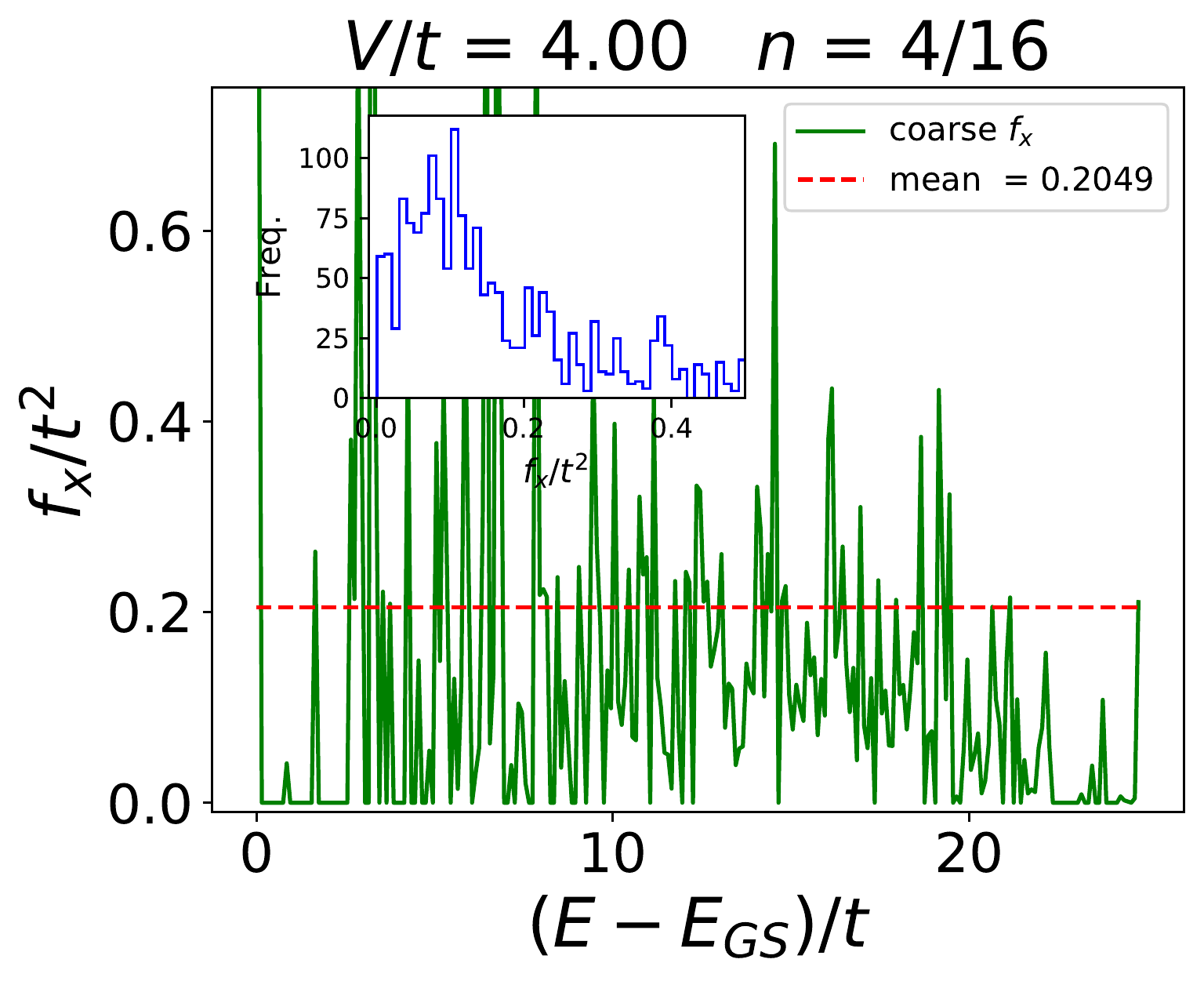}
\includegraphics[width=0.33\linewidth]{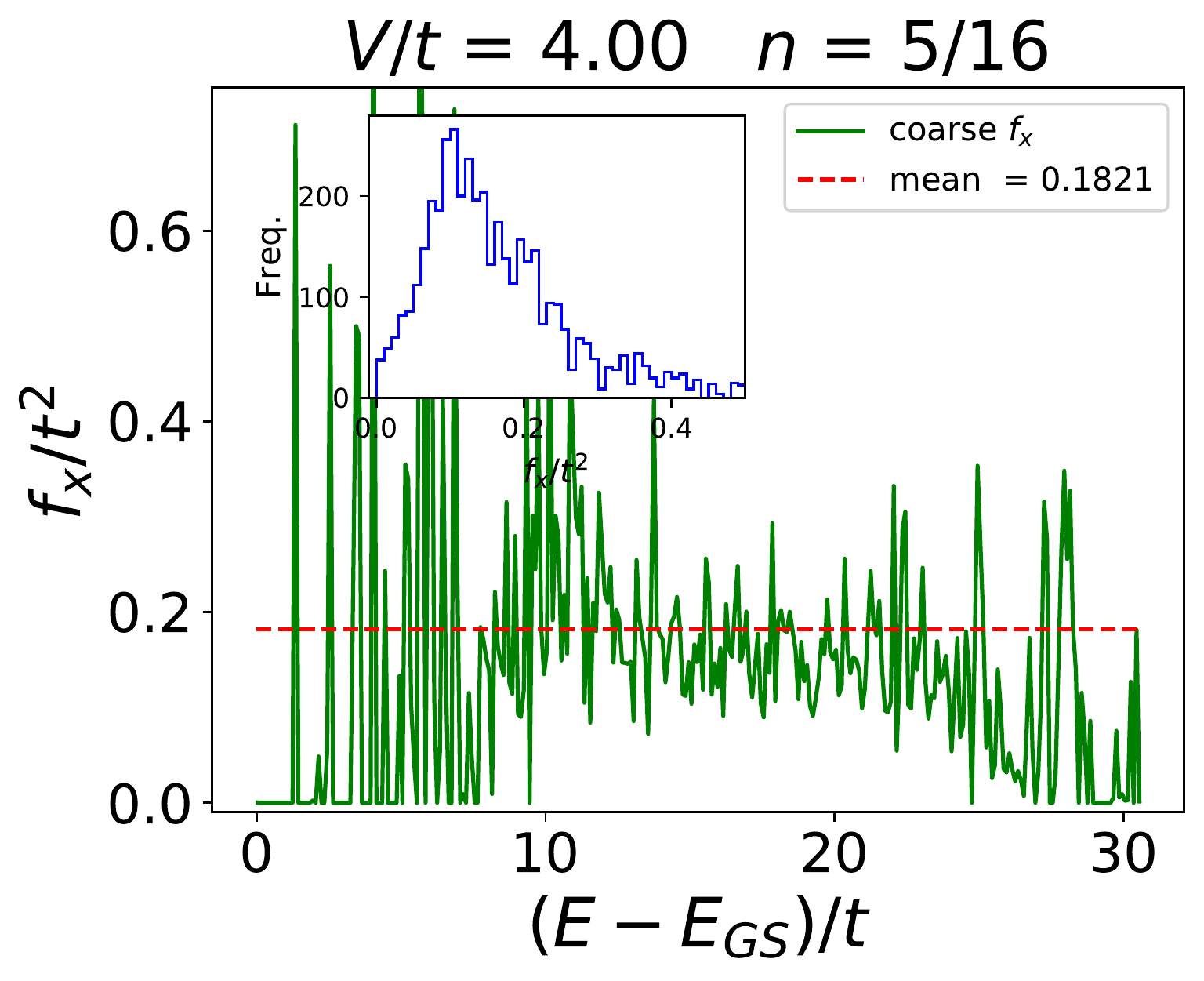}
\includegraphics[width=0.33\linewidth]{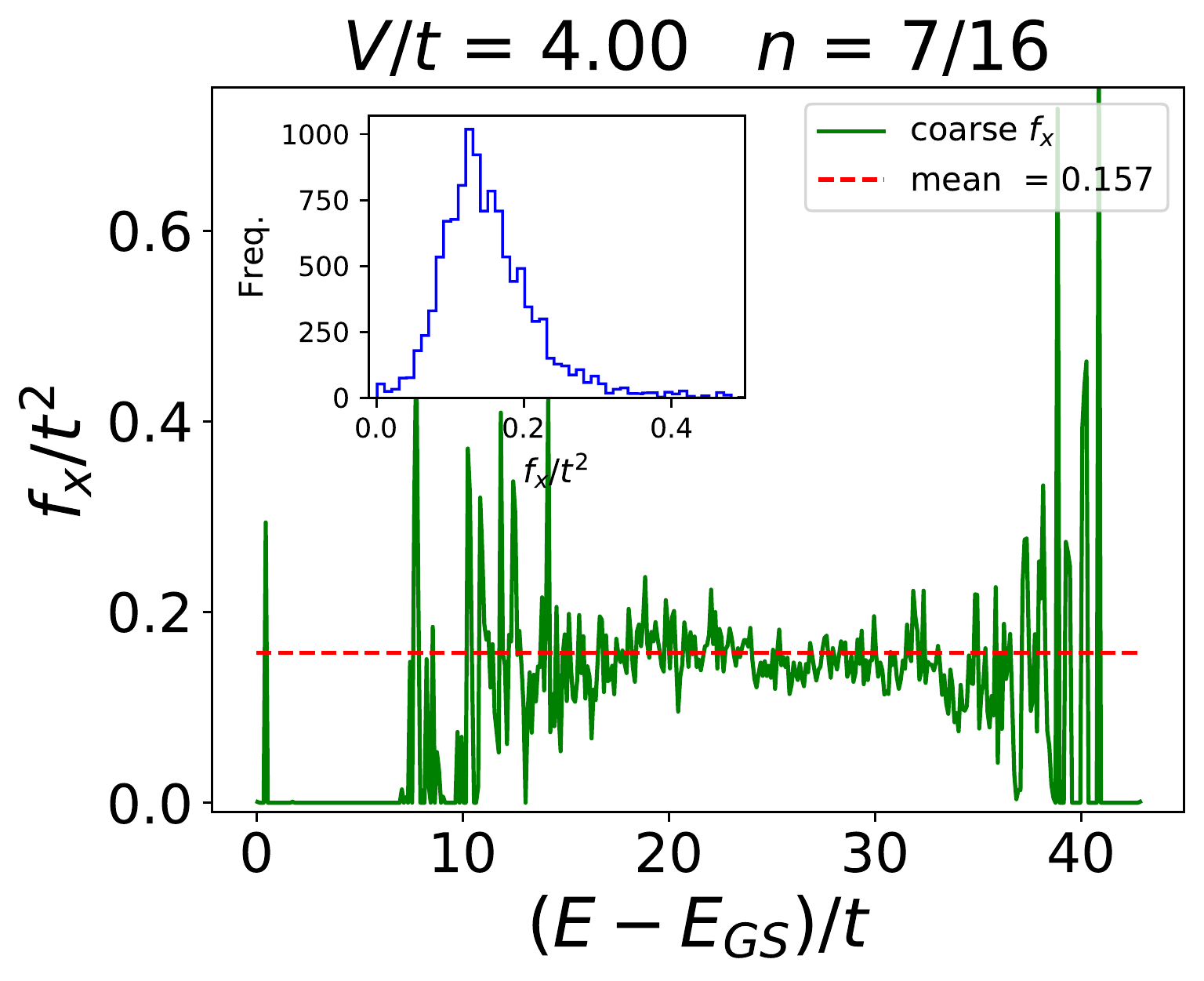}
\includegraphics[width=0.33\linewidth]{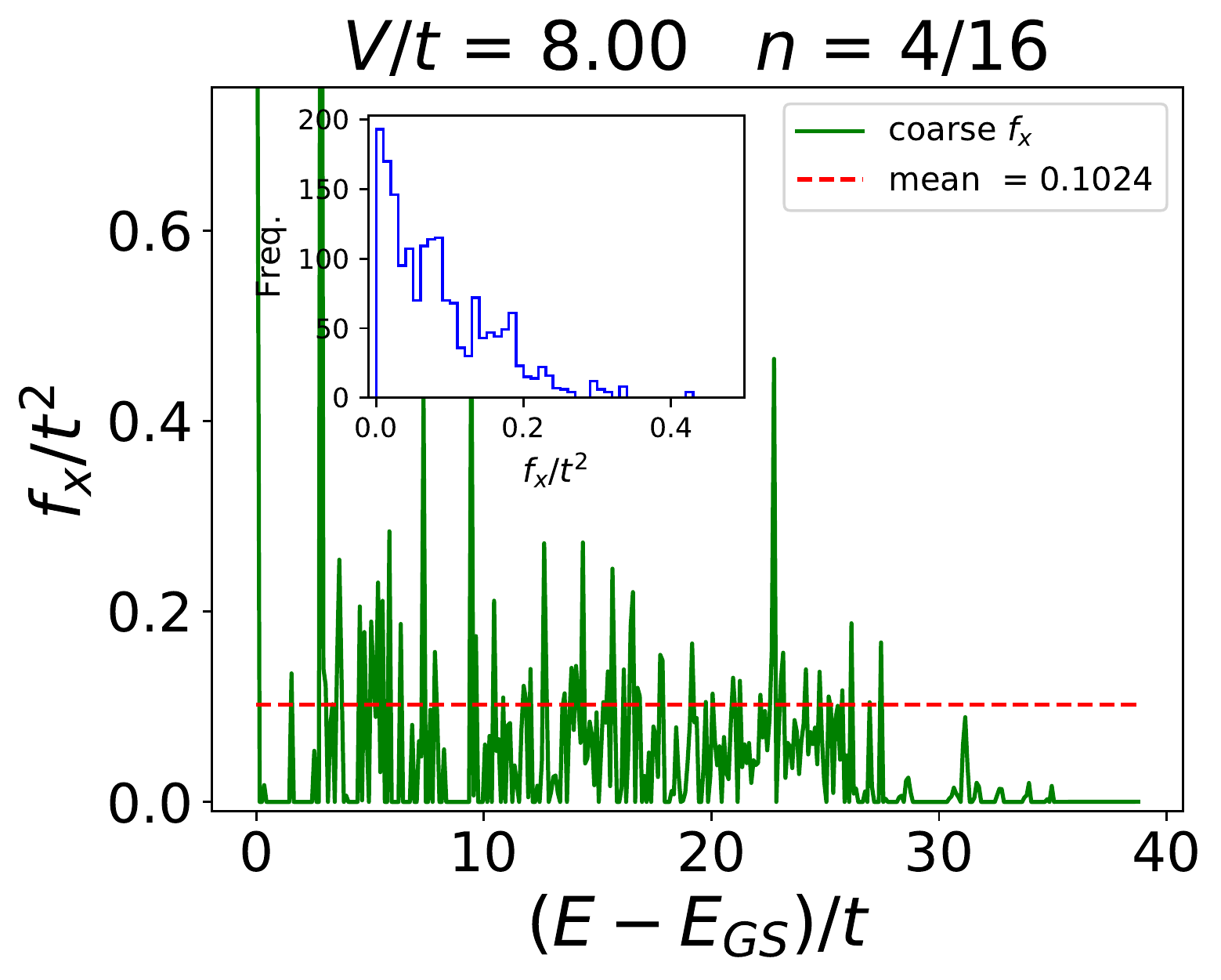}
\includegraphics[width=0.33\linewidth]{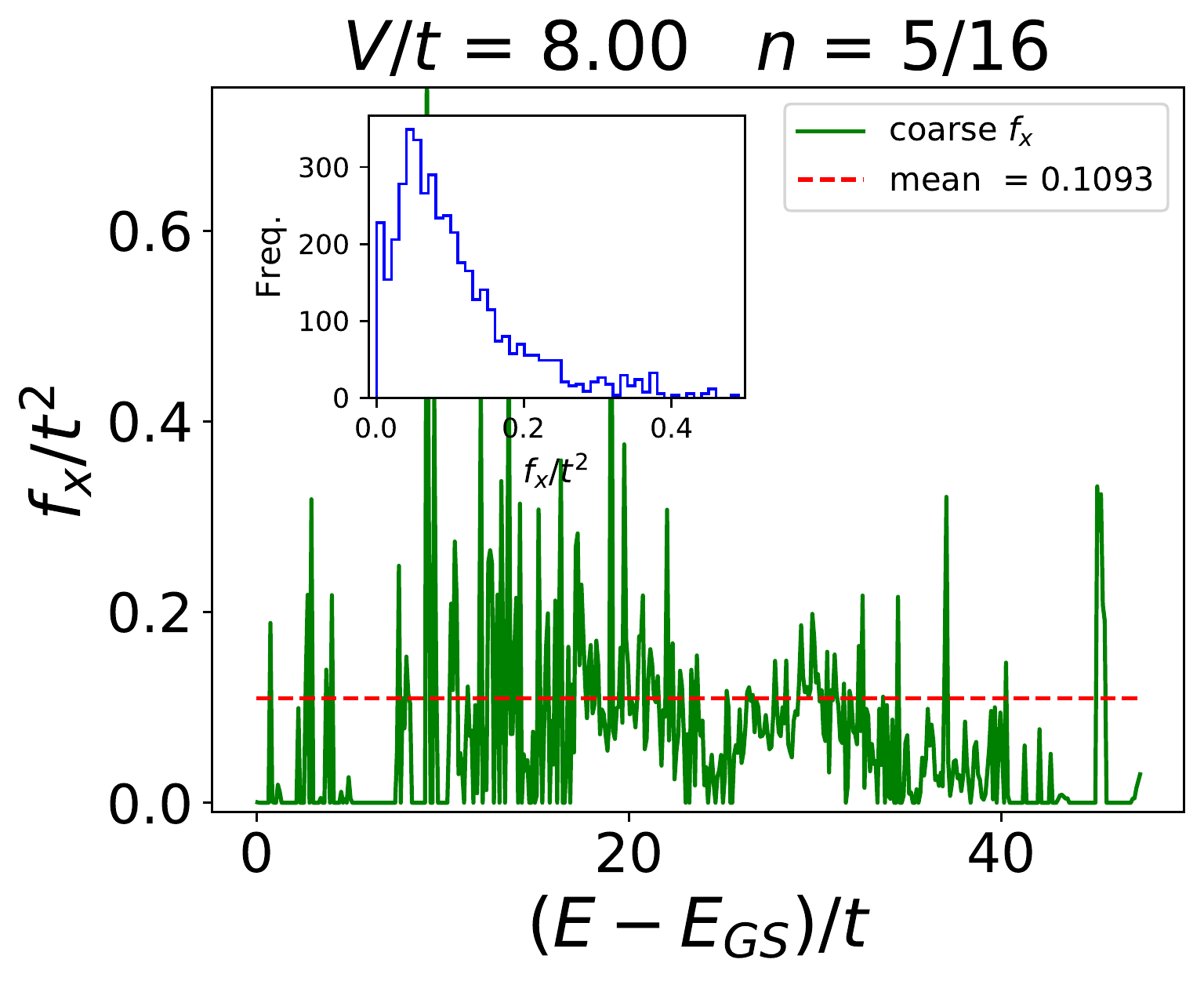}
\includegraphics[width=0.33\linewidth]{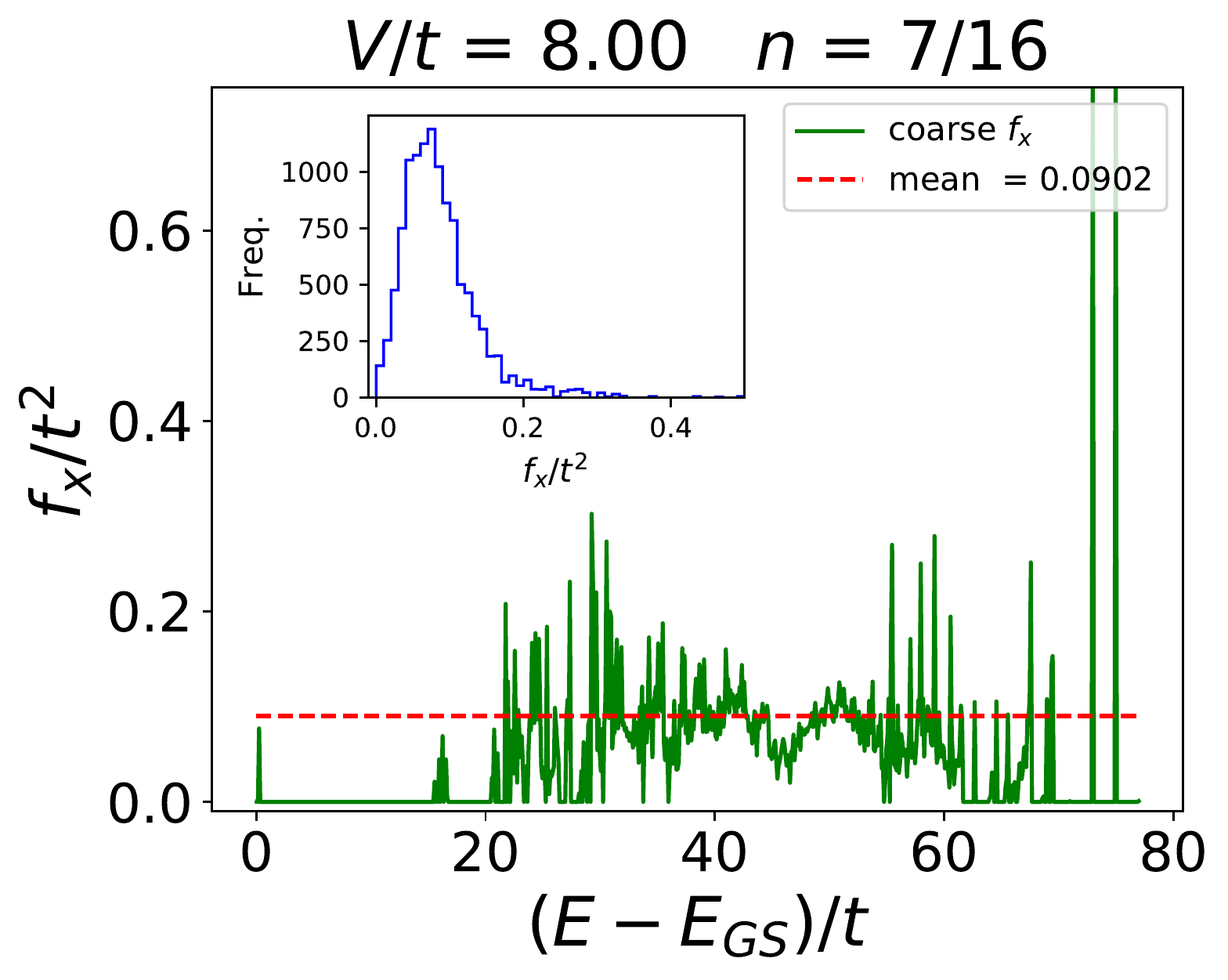}
\includegraphics[width=0.33\linewidth]{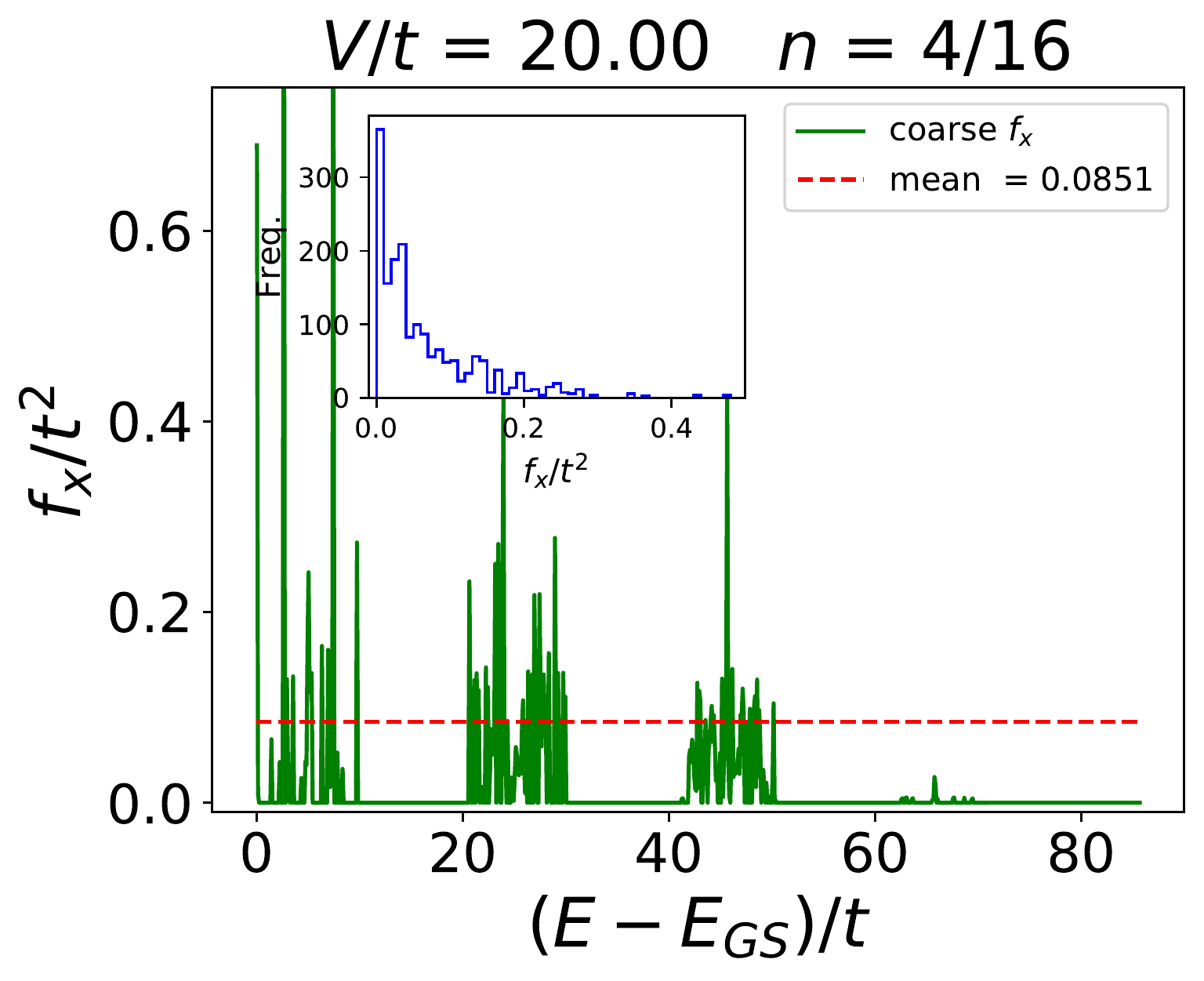}
\includegraphics[width=0.33\linewidth]{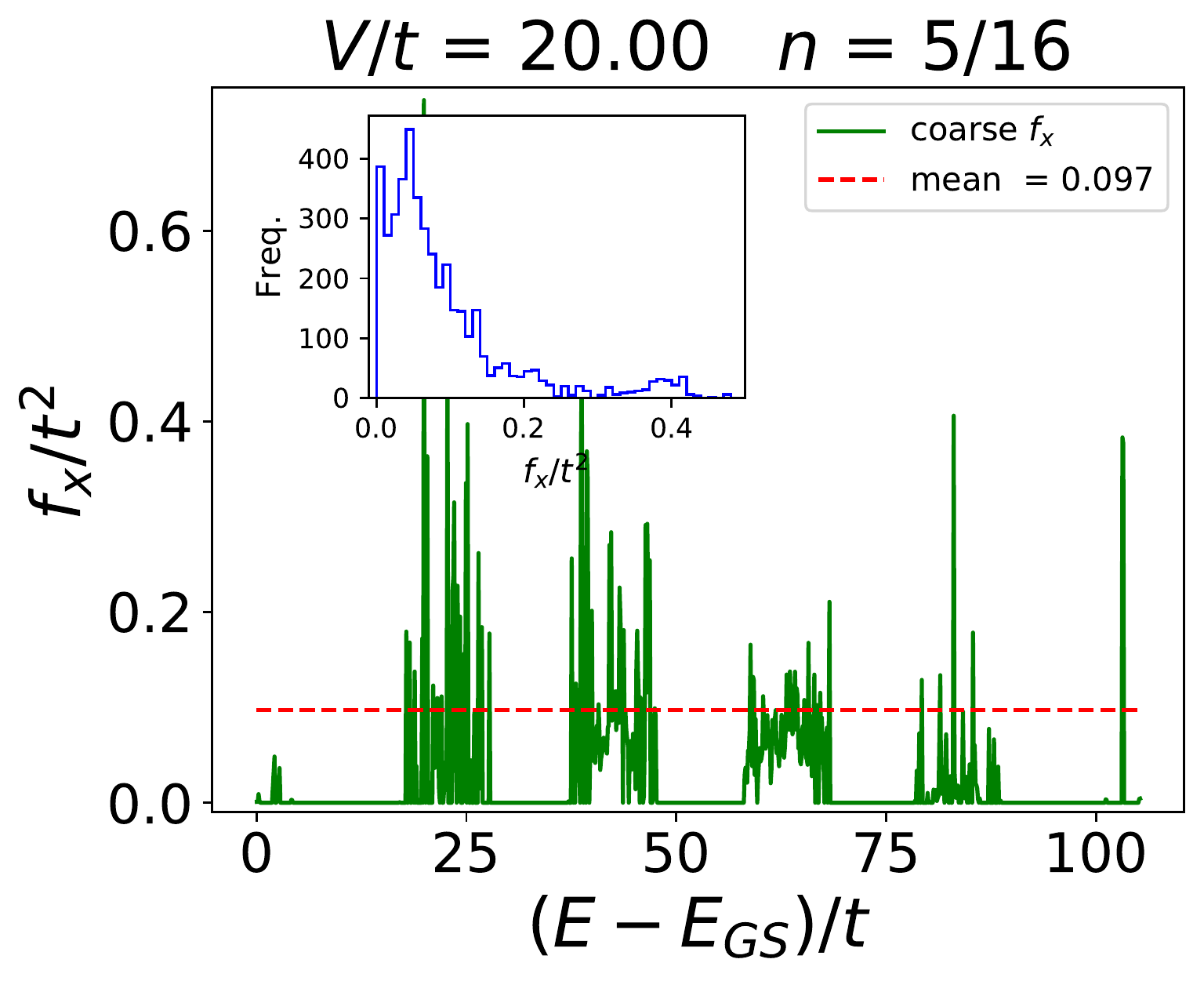}
\includegraphics[width=0.33\linewidth]{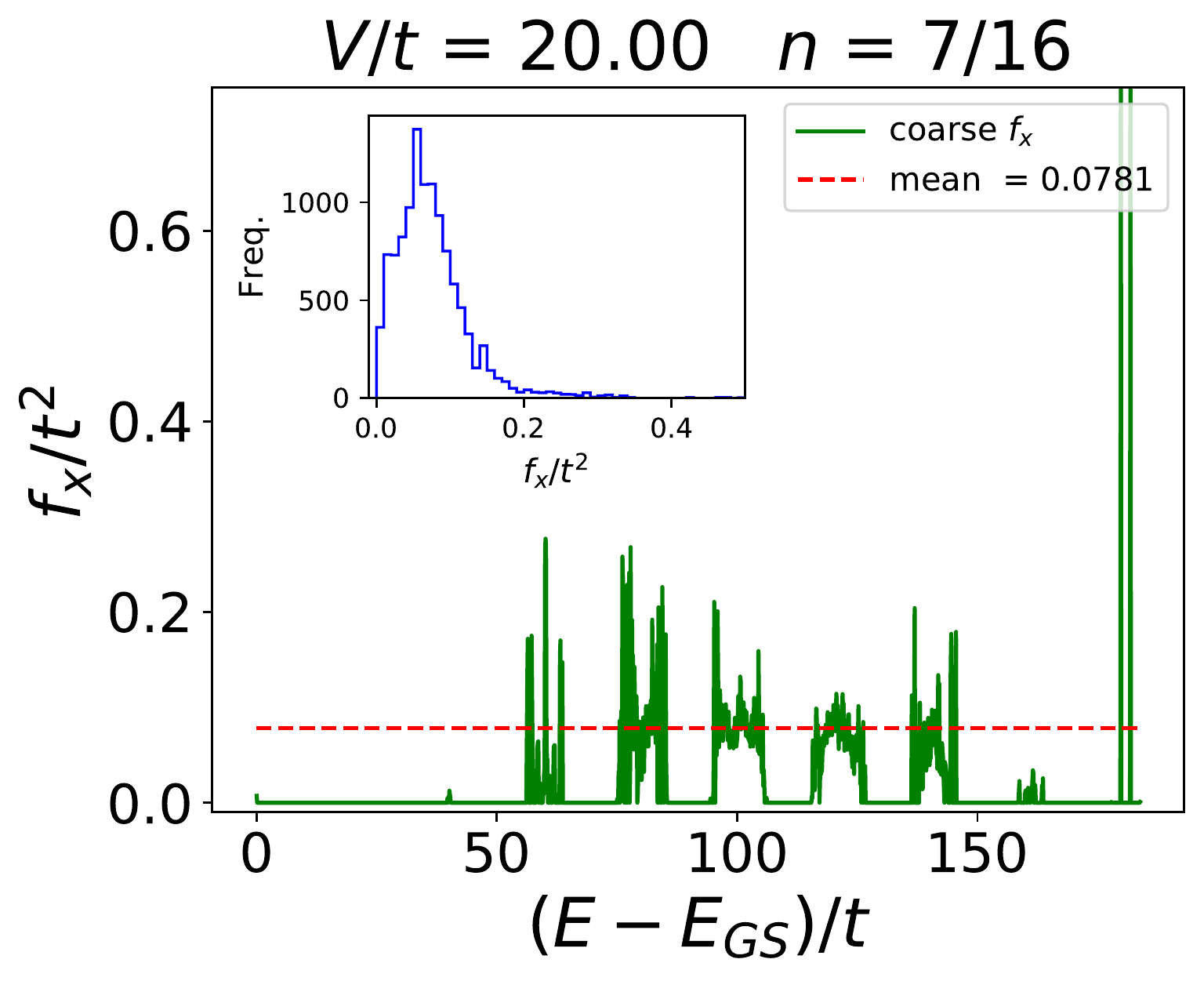}
\caption{$f_x(E)$ for the charge conductivity, using $\eta=0.10 t$ as a function of the energy $E$ for the $4 \times 4$ 2D square lattice nearest neighbor spinless Hubbard model for representative fillings for $V/t = 4,8,20$. In each case, the ground state energy $E_{GS}$ has been subtracted out. In the last row of figures, $f_x(E)$ vanishes in between the regions where it is roughly flat, because the spectrum itself does not have support within these energy windows. This is consistent with the requirements for our proof of $f$-invariance, which just needs $f$ to take the same value on all regions where the spectrum has support, and therefore where $g(E)\neq0$. The insets show histograms of $f_x(E)$ values with the bin width set to 0.01.}
\label{fig:ffn_spinless_Hubbard_1}
\end{figure}

\begin{figure}
\includegraphics[width=0.33\linewidth]{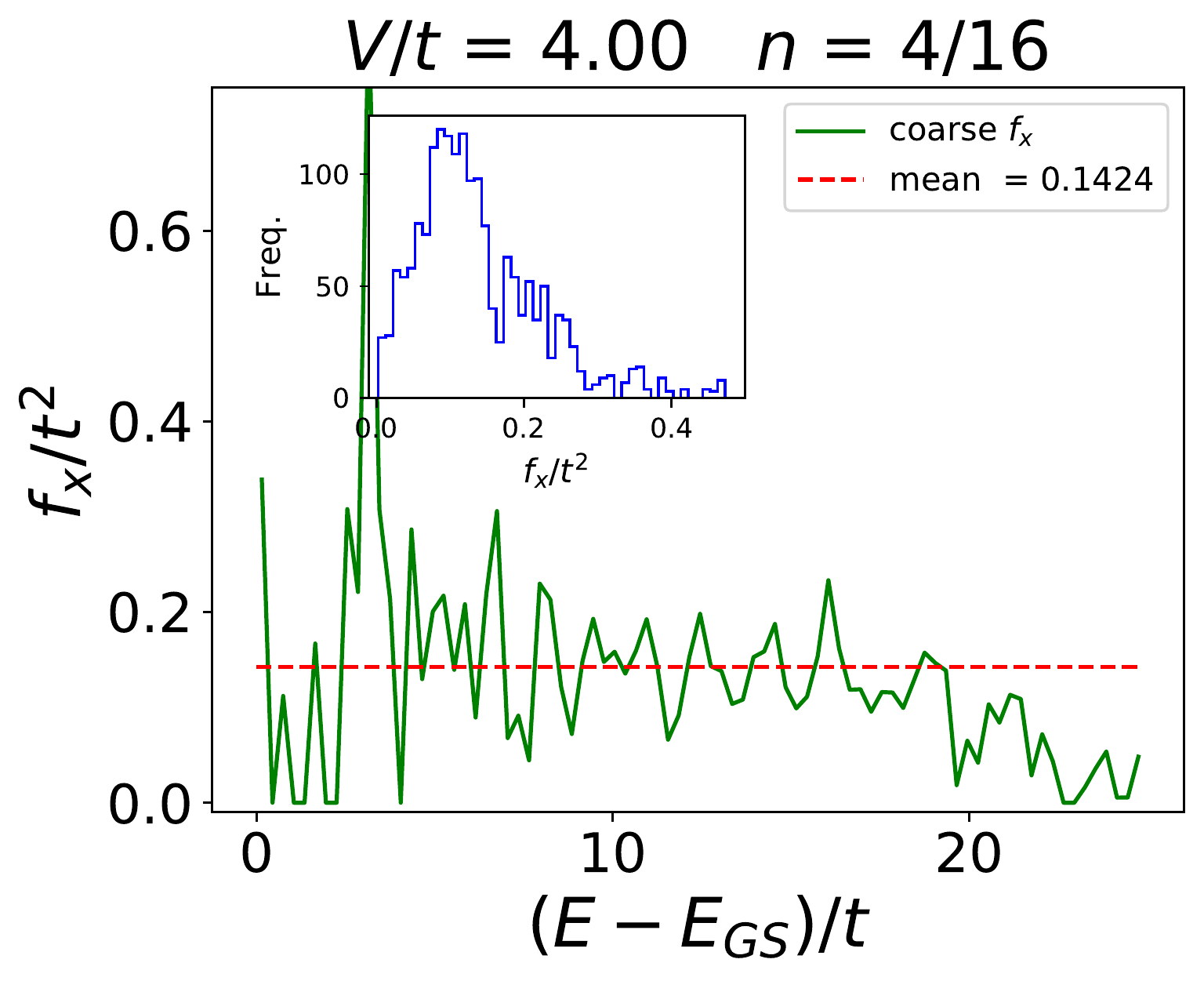}
\includegraphics[width=0.33\linewidth]{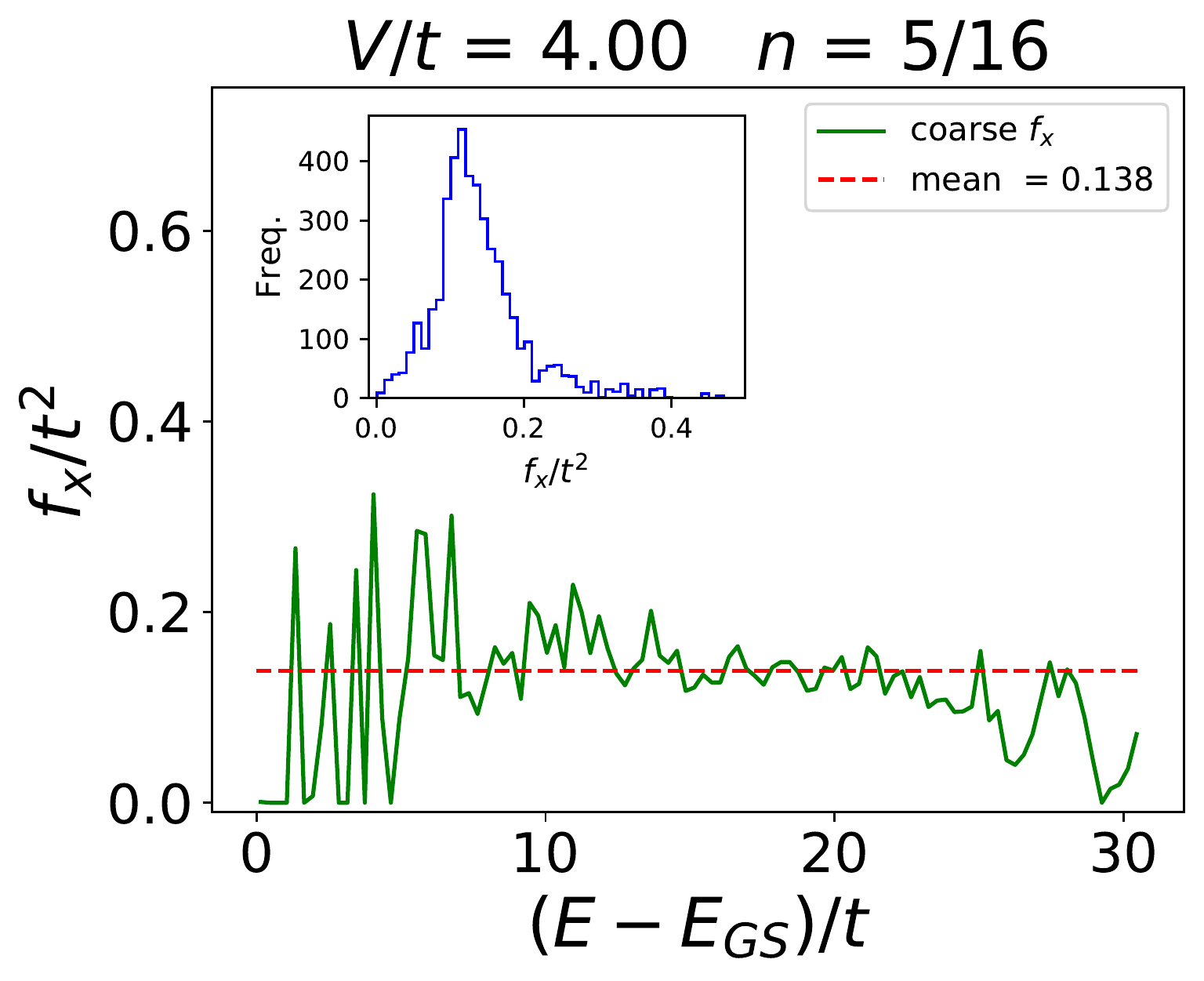}
\includegraphics[width=0.33\linewidth]{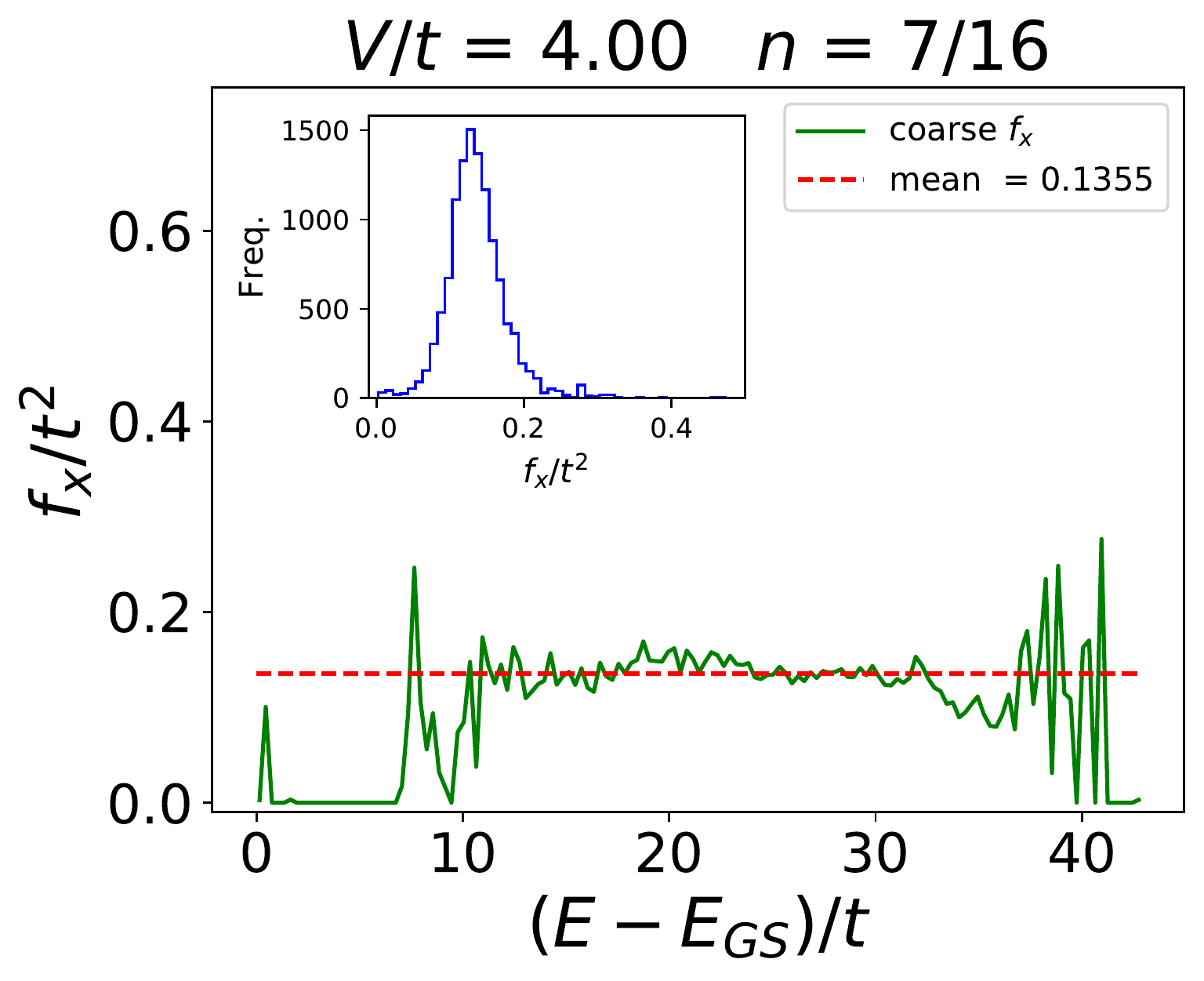}
\includegraphics[width=0.33\linewidth]{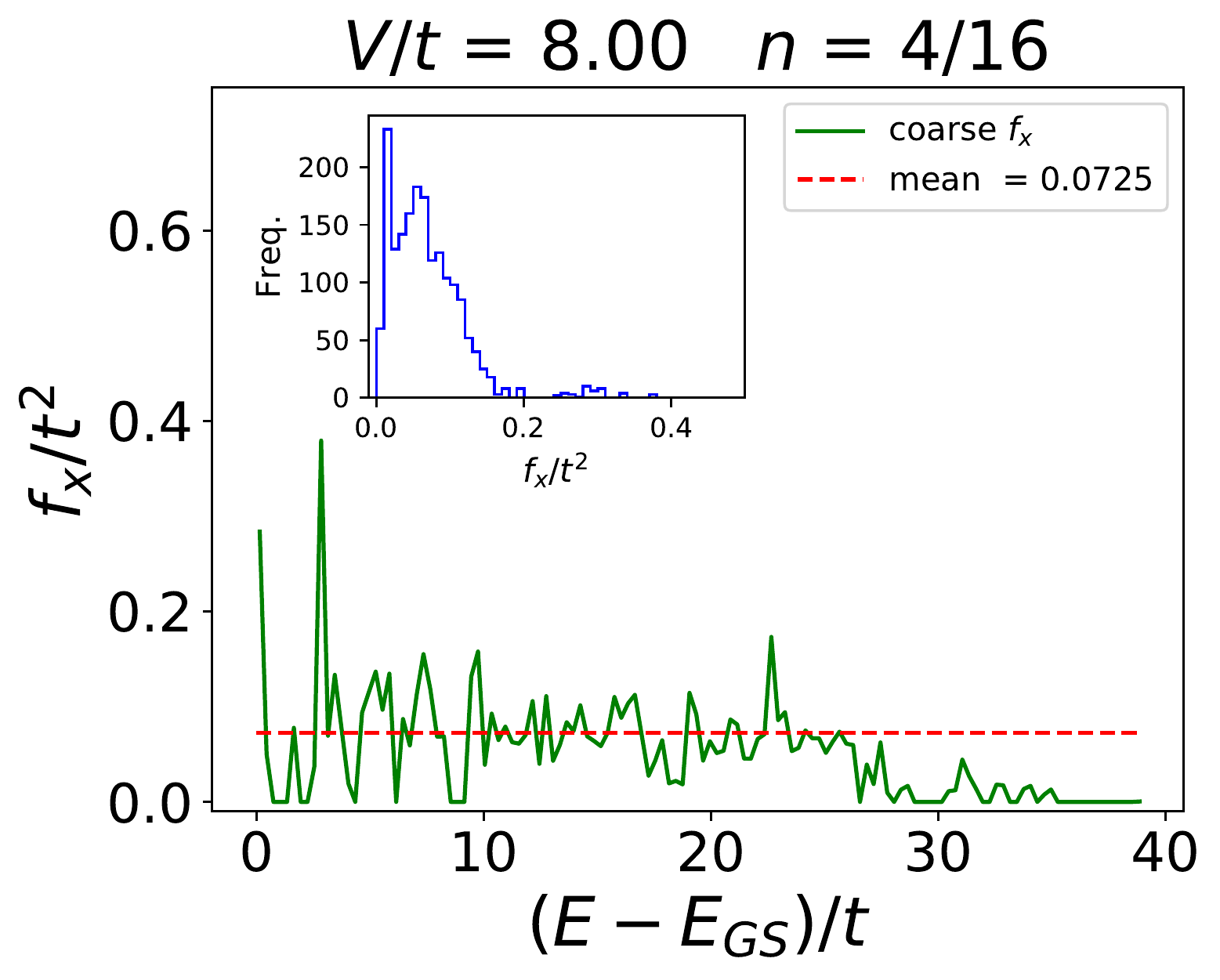}
\includegraphics[width=0.33\linewidth]{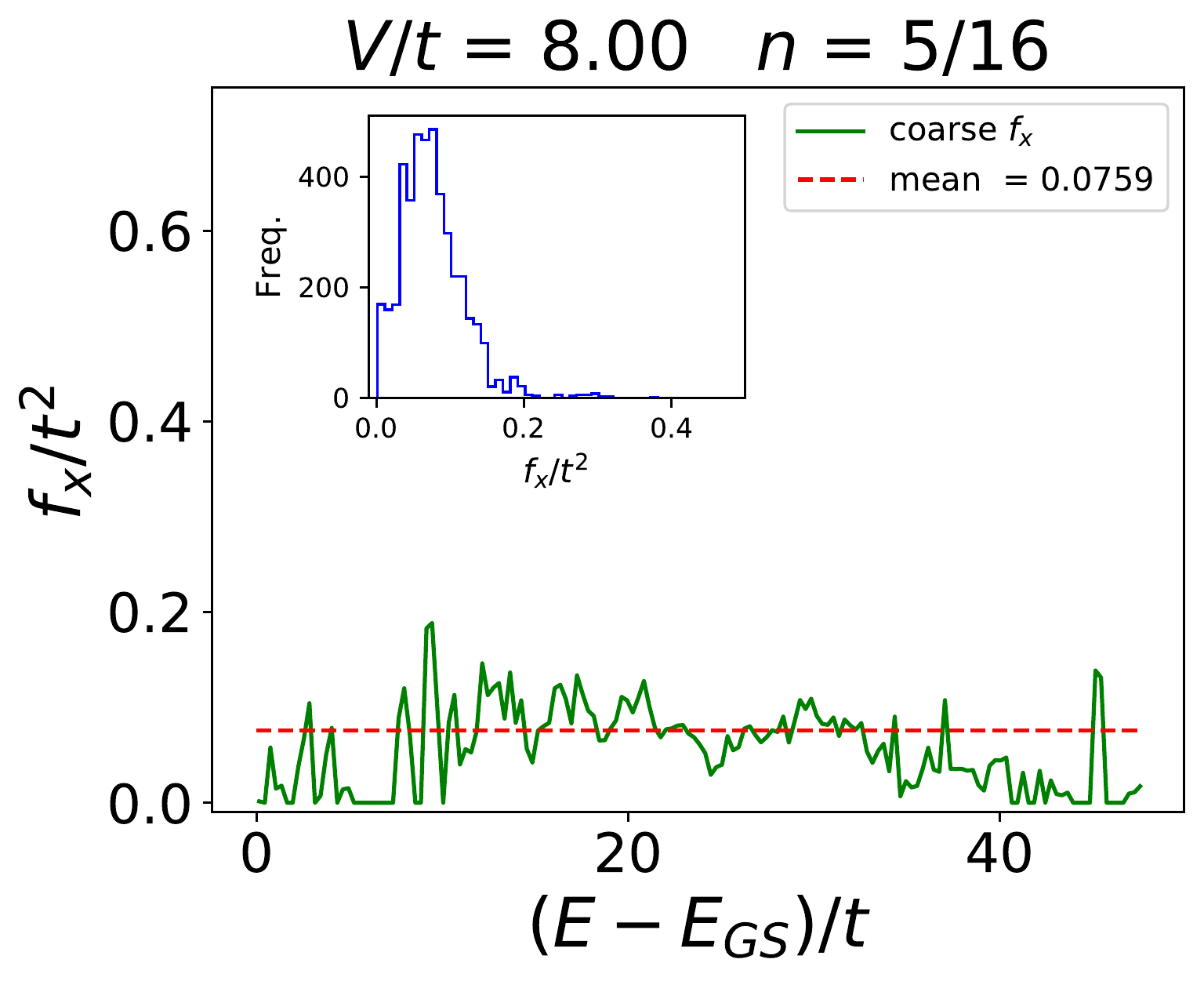}
\includegraphics[width=0.33\linewidth]{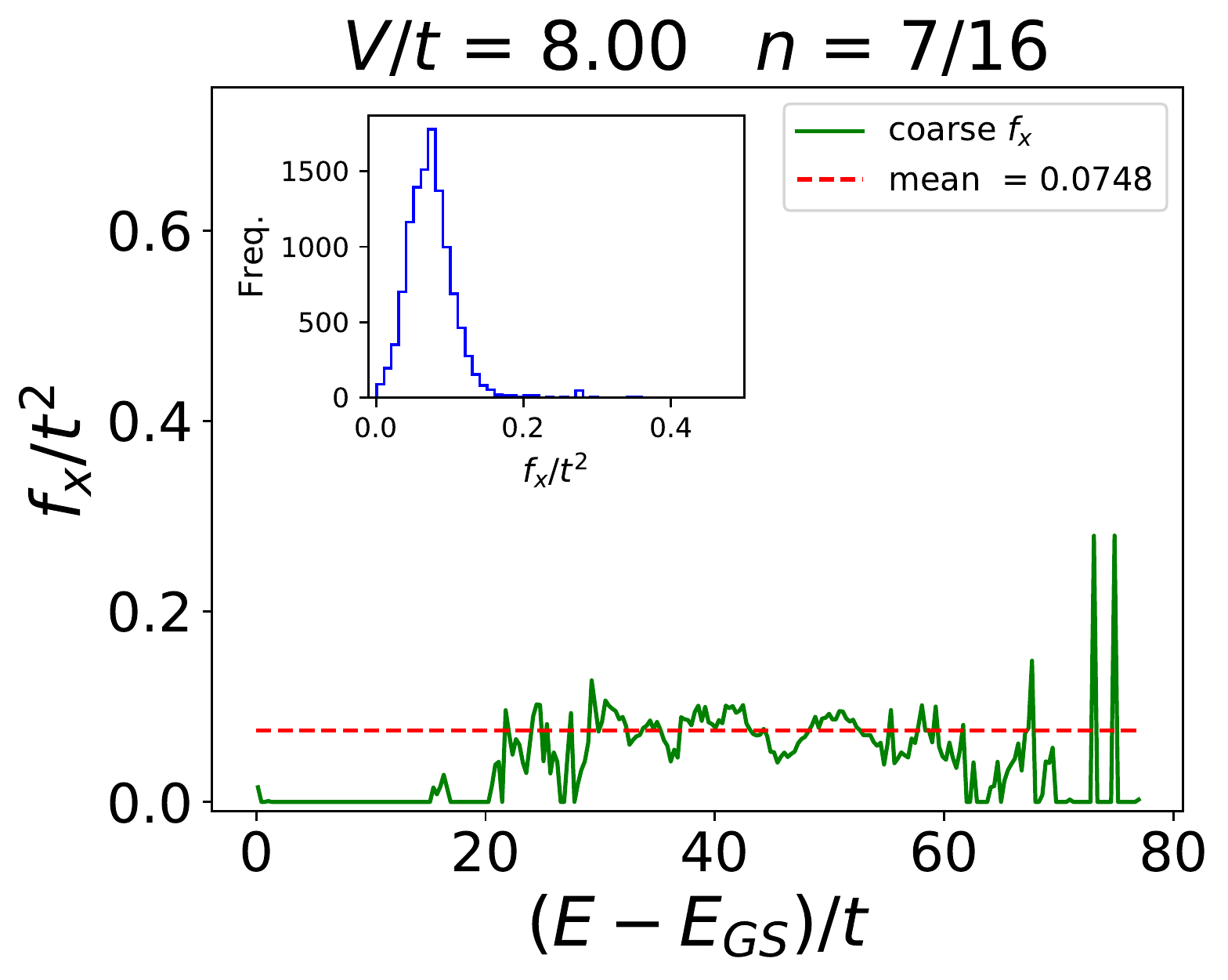}
\includegraphics[width=0.33\linewidth]{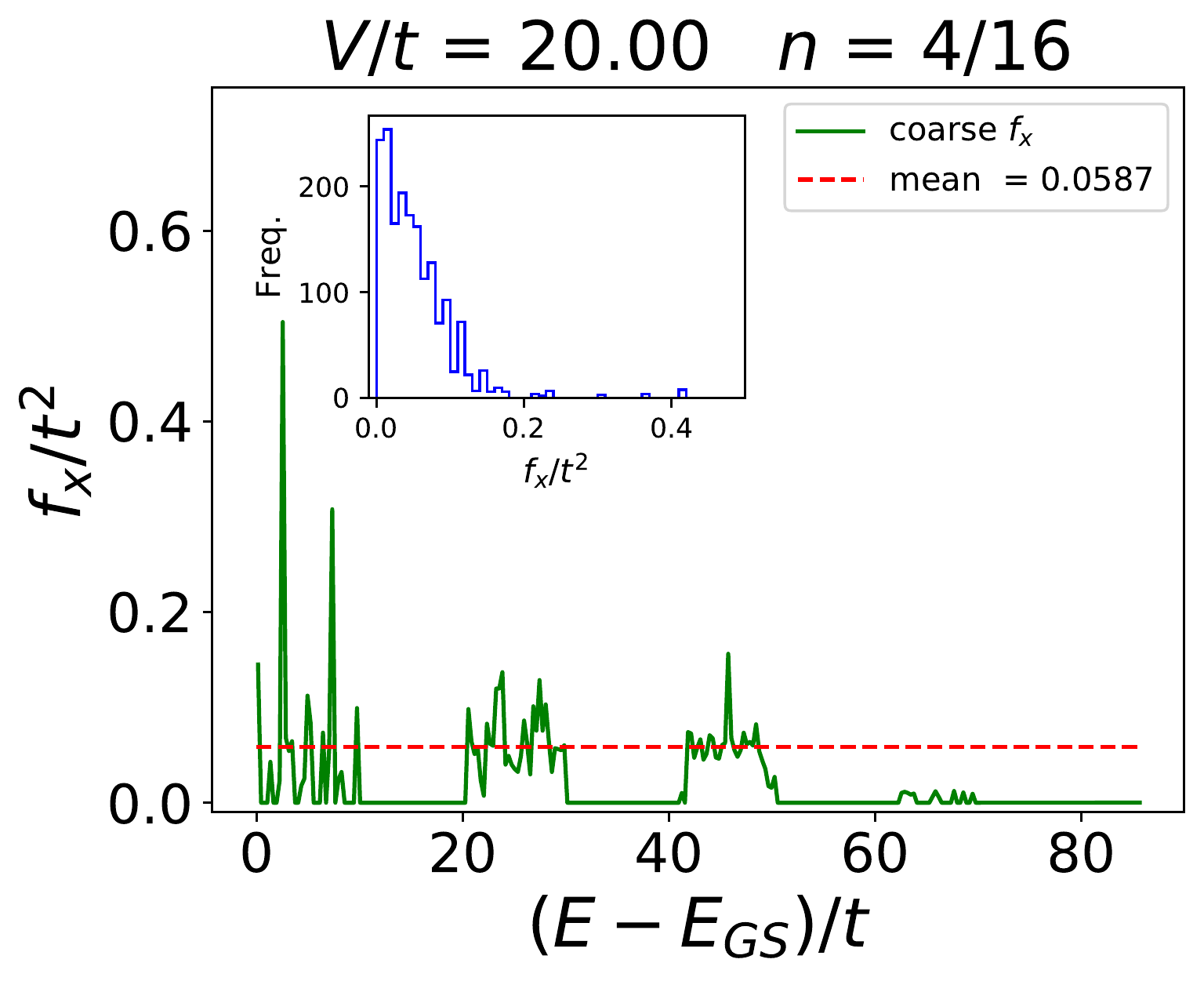}
\includegraphics[width=0.33\linewidth]{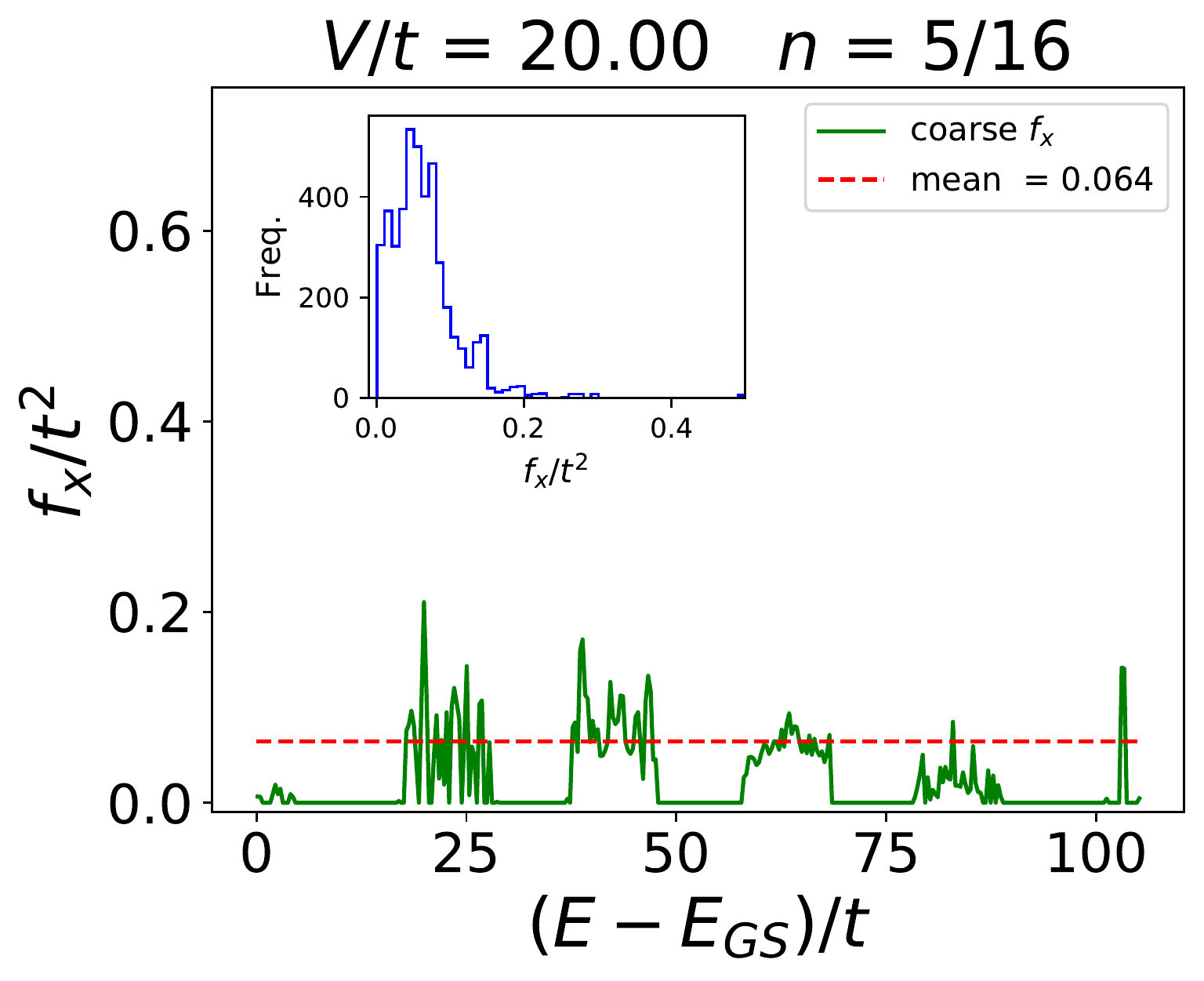}
\includegraphics[width=0.33\linewidth]{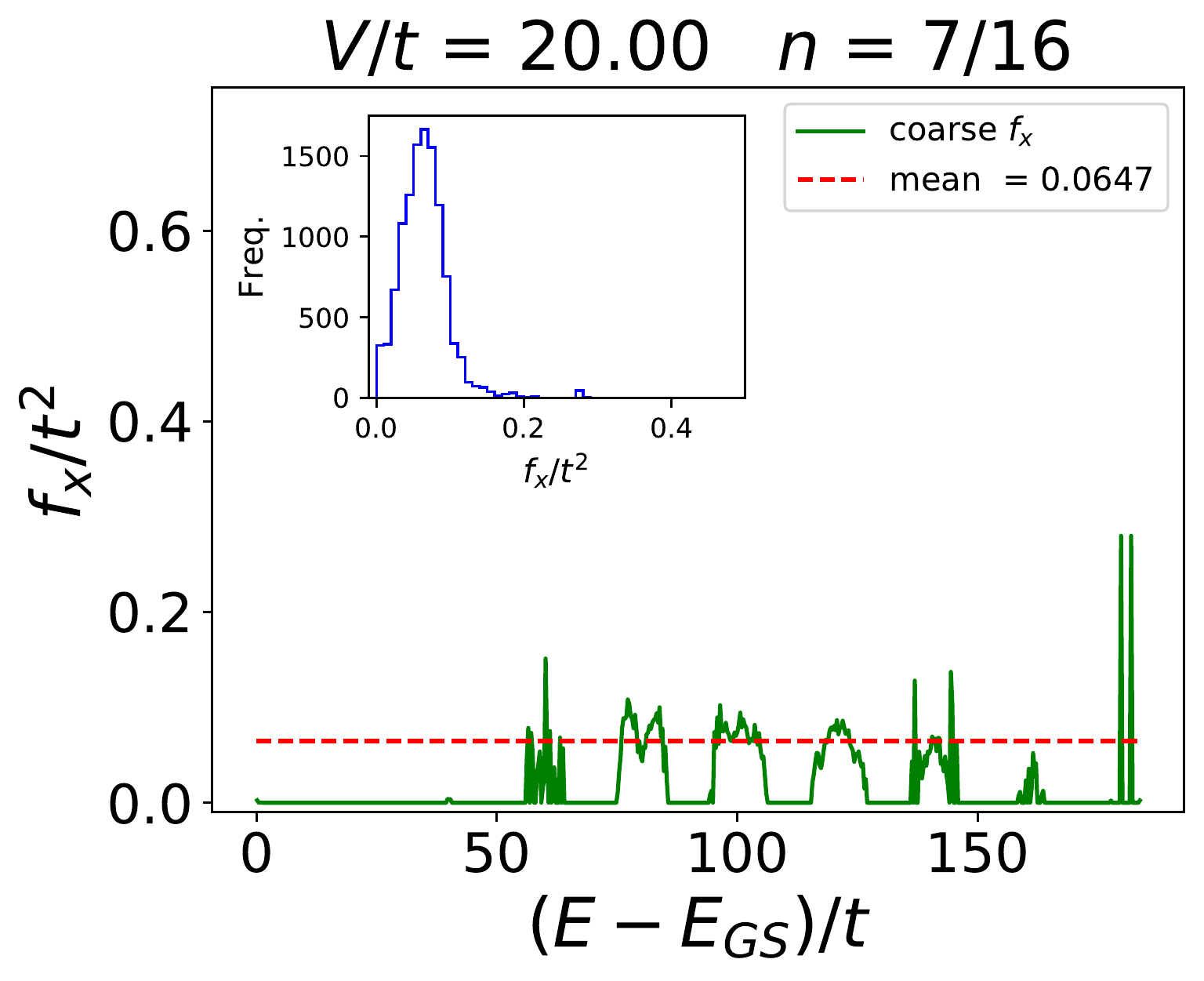}
\caption{$f_x(E)$ for the charge conductivity, using $\eta=0.30 t$}
\label{fig:ffn_spinless_Hubbard_2}
\end{figure}

\begin{figure}
\includegraphics[width=0.33\linewidth]{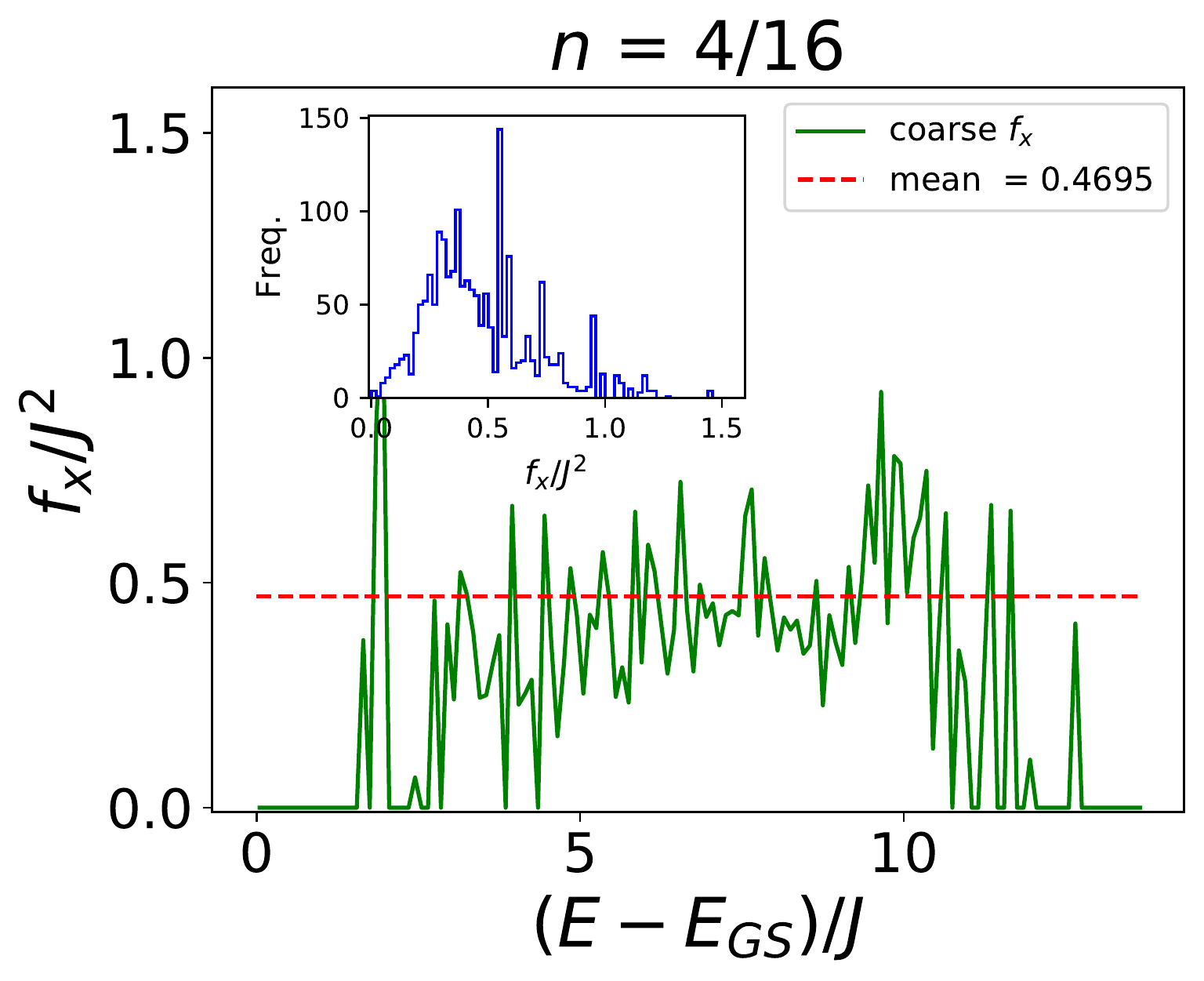}
\includegraphics[width=0.33\linewidth]{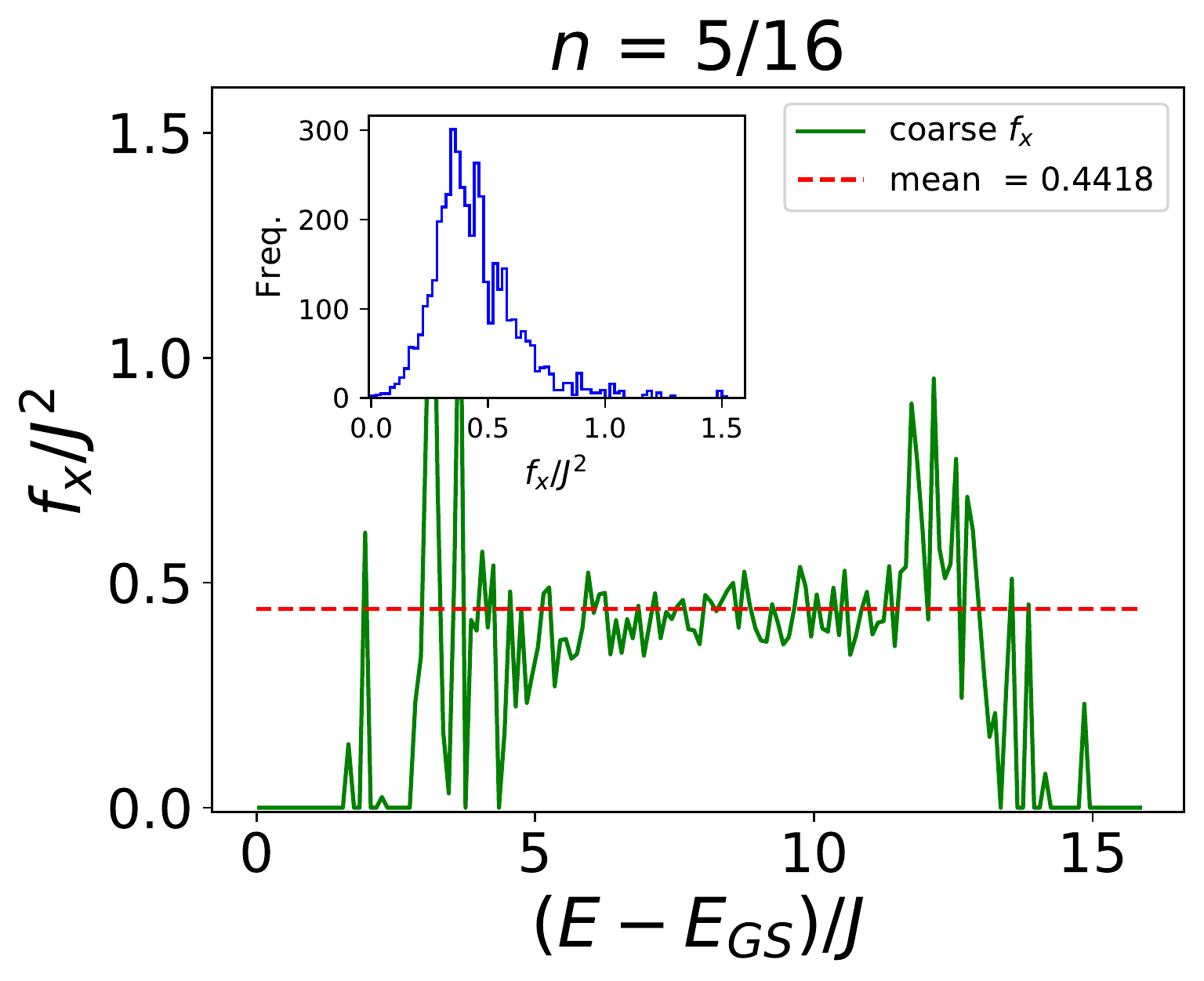}
\includegraphics[width=0.33\linewidth]{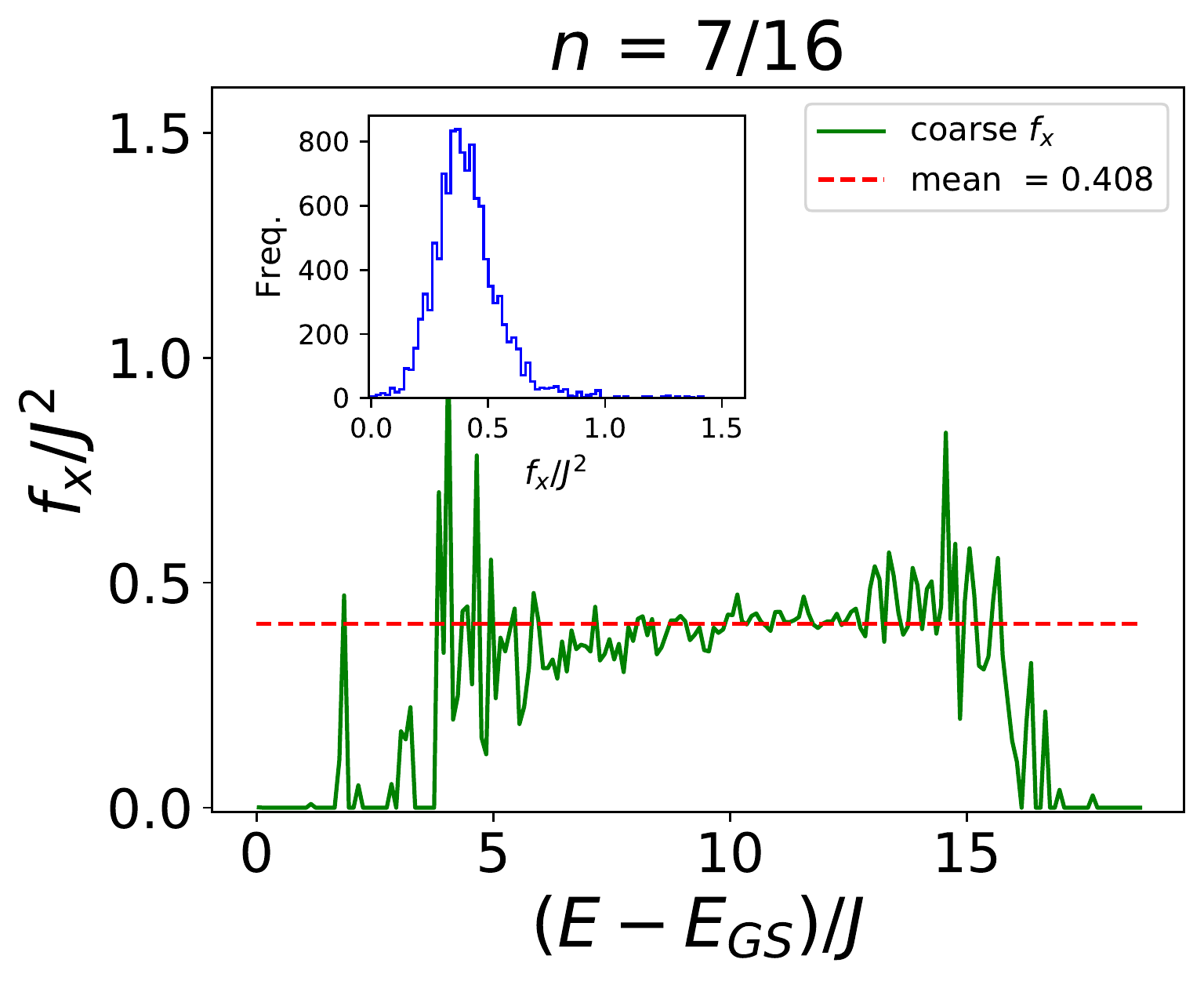}
\includegraphics[width=0.33\linewidth]{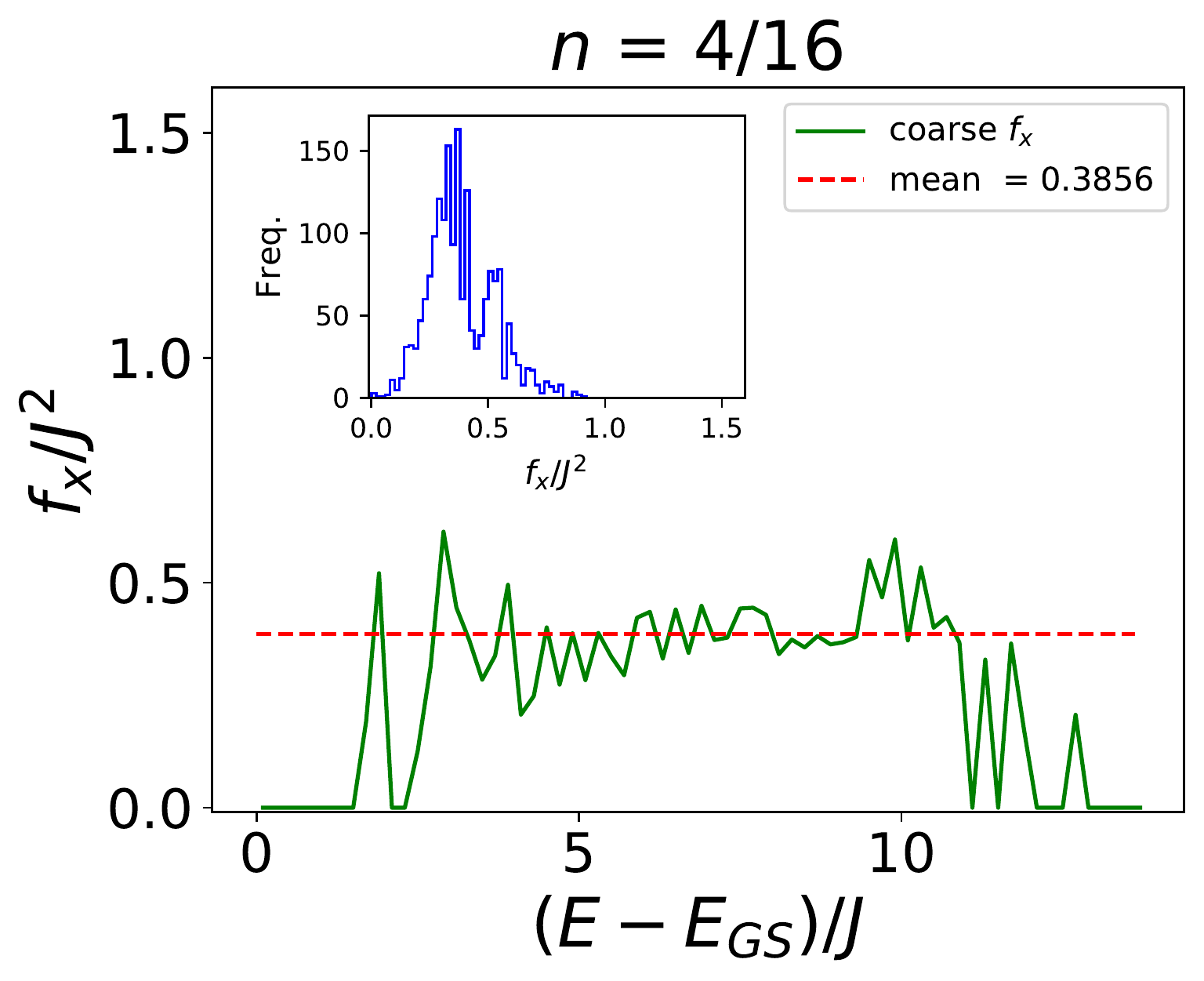}
\includegraphics[width=0.33\linewidth]{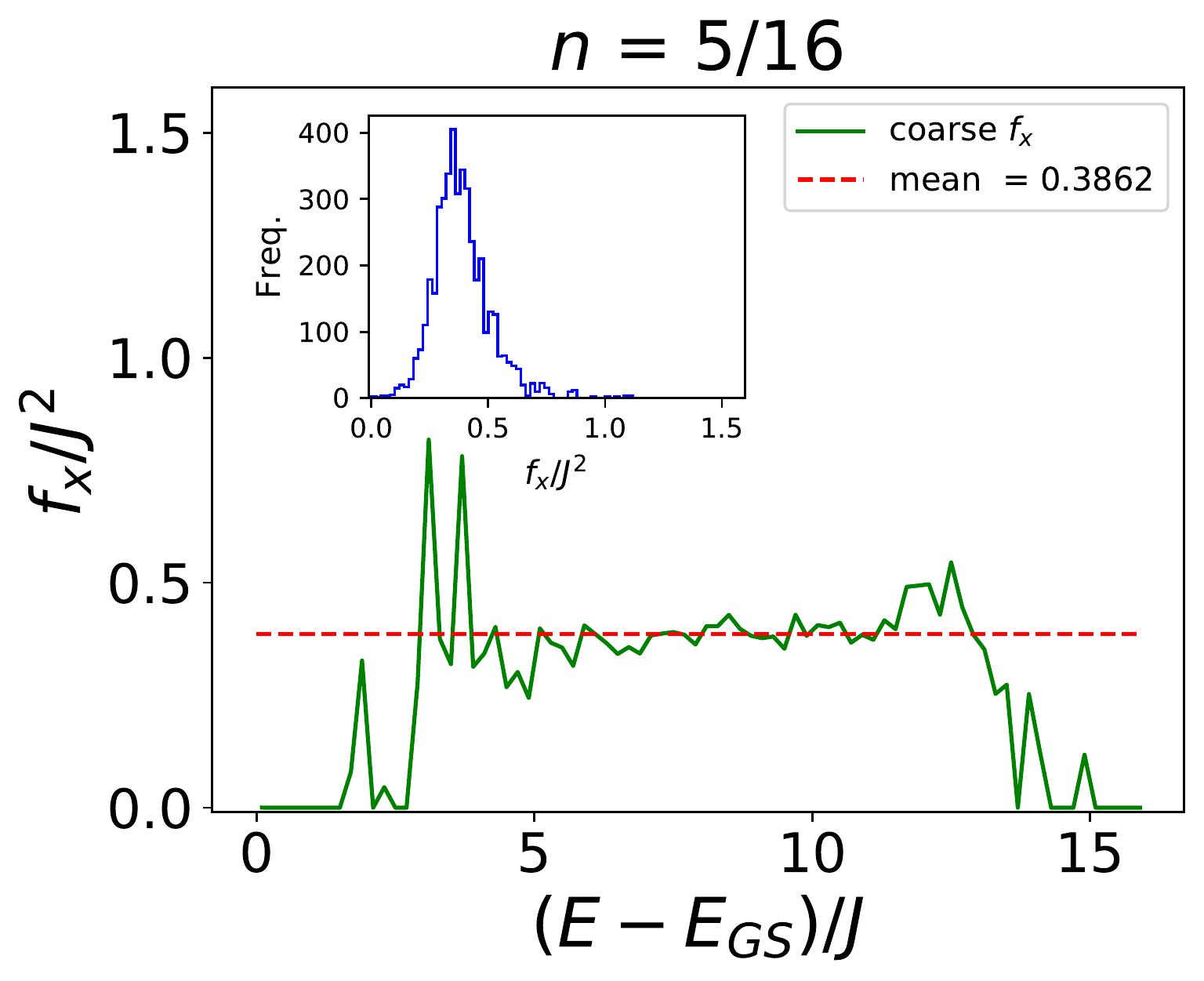}
\includegraphics[width=0.33\linewidth]{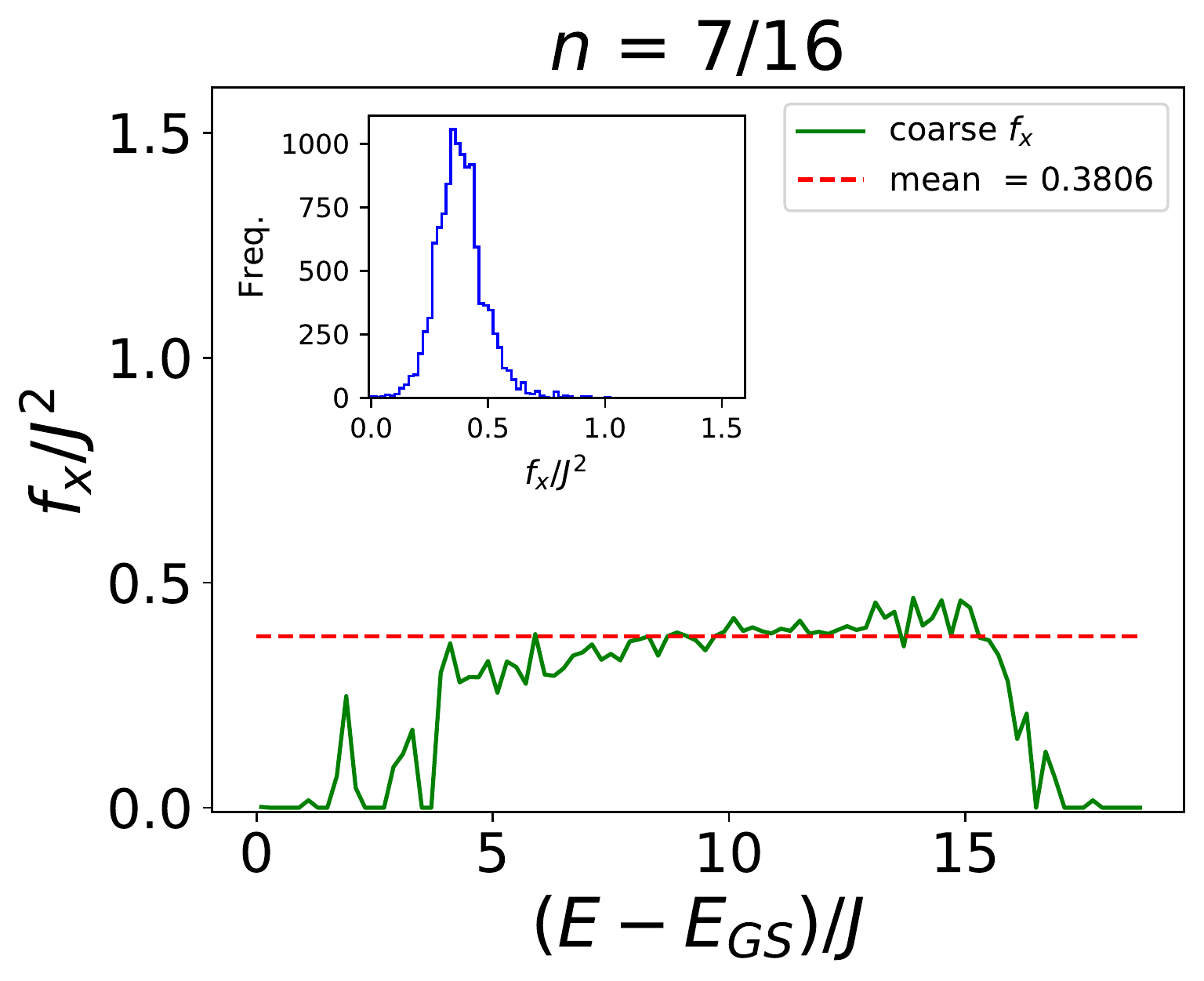}
\includegraphics[width=0.33\linewidth]{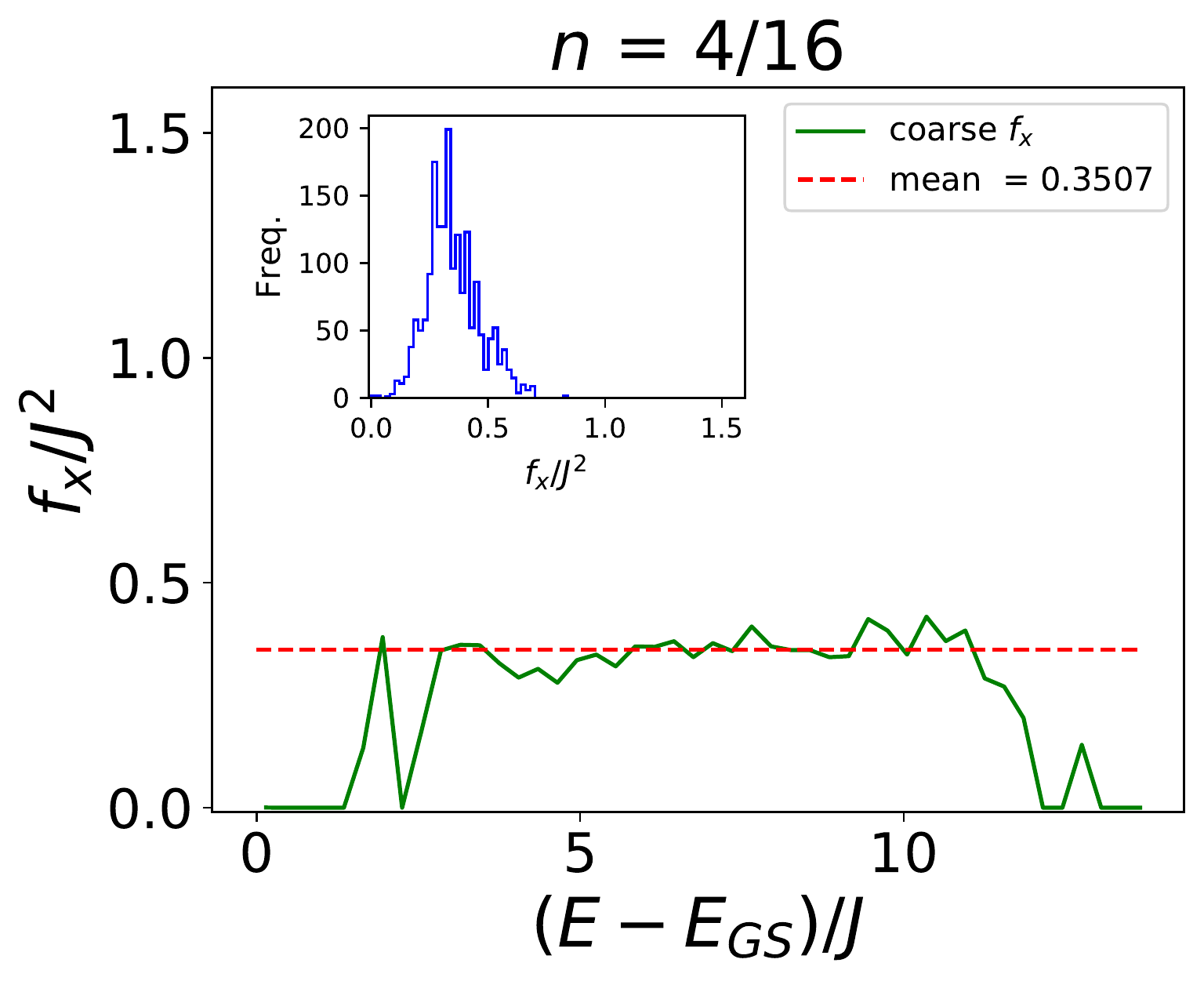}
\includegraphics[width=0.33\linewidth]{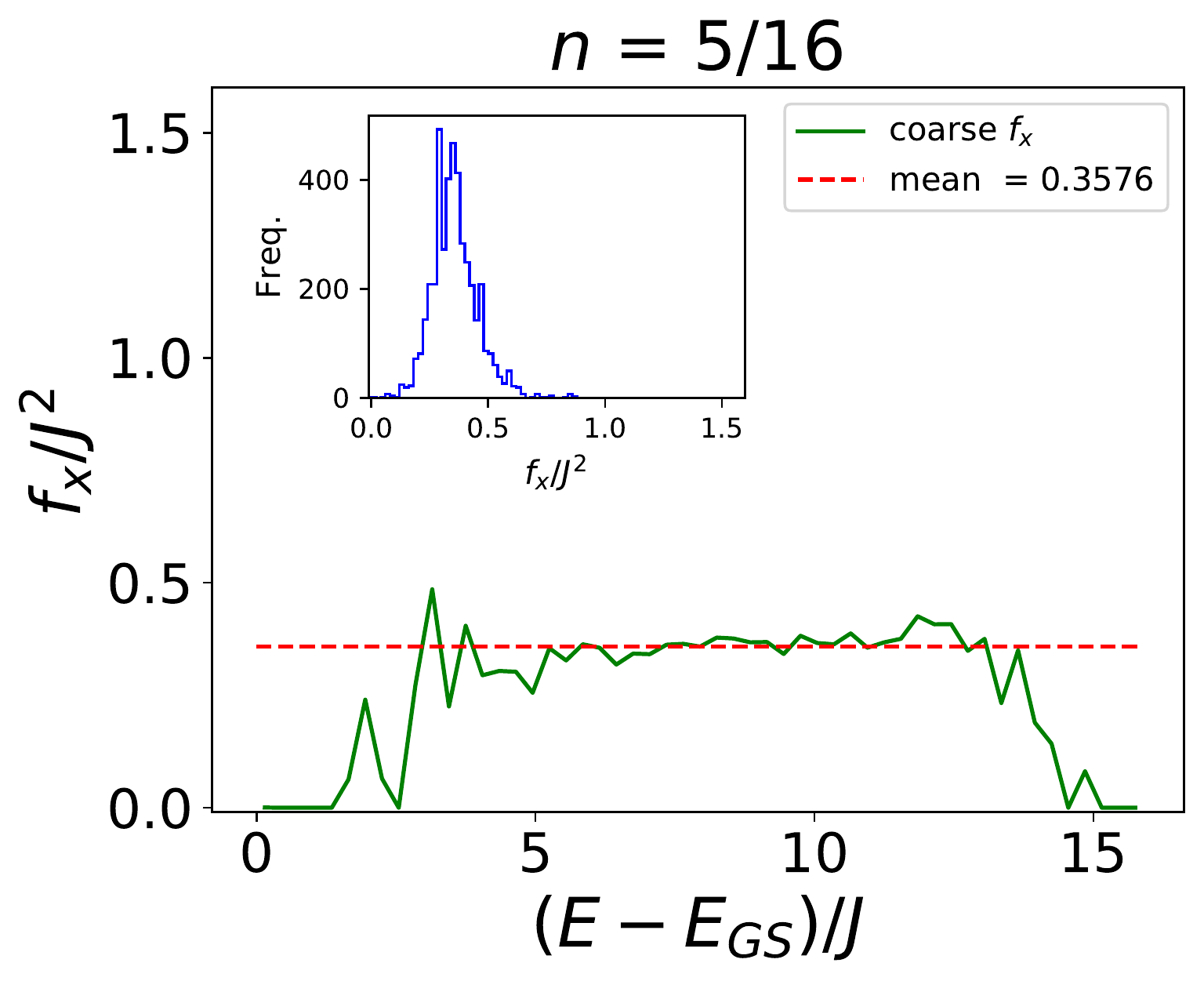}
\includegraphics[width=0.33\linewidth]{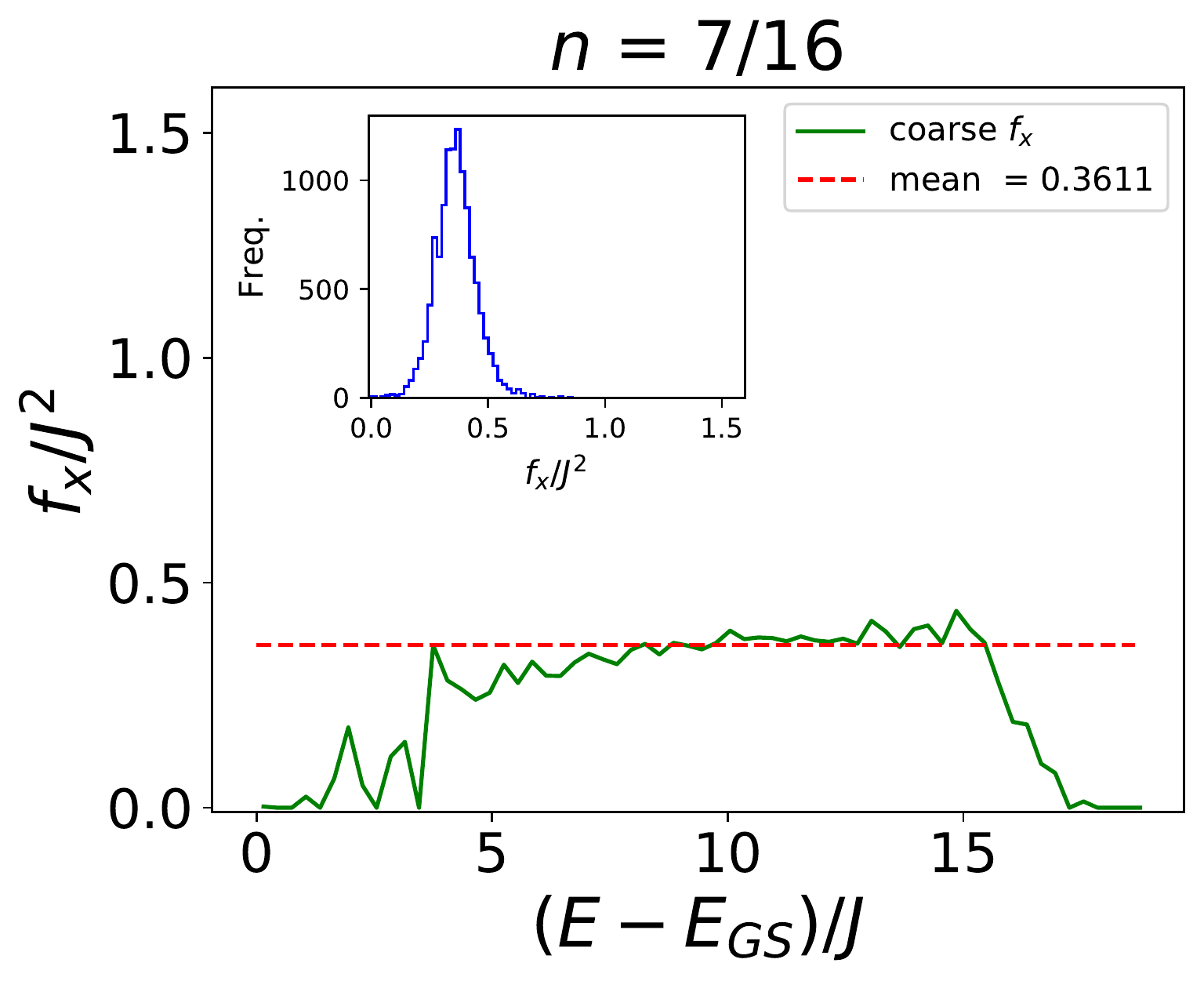}
\caption{$f_x(E)$ for the spin conductivity of the $4\times 4$ 2D square lattice nearest neighbor spin 1/2 Heisenberg model for representative magnetizations (equivalently fillings of spin-up particles in a sea of down spins, denoted by $n$) using the Lorentzian protocol with broadening parameter $\eta$; $\eta=0.10 J$ (top row), $\eta=0.20 J$ (middle row) and $\eta = 0.30 J$ (bottom row). In each case, the ground state energy $E_{GS}$ has been subtracted out. The insets show histograms of $f_x(E)$ values with the bin width set to 0.02.}
\label{fig:ffn_Heisenberg_1}
\end{figure} 

\begin{figure}
\includegraphics[width=0.33\linewidth]{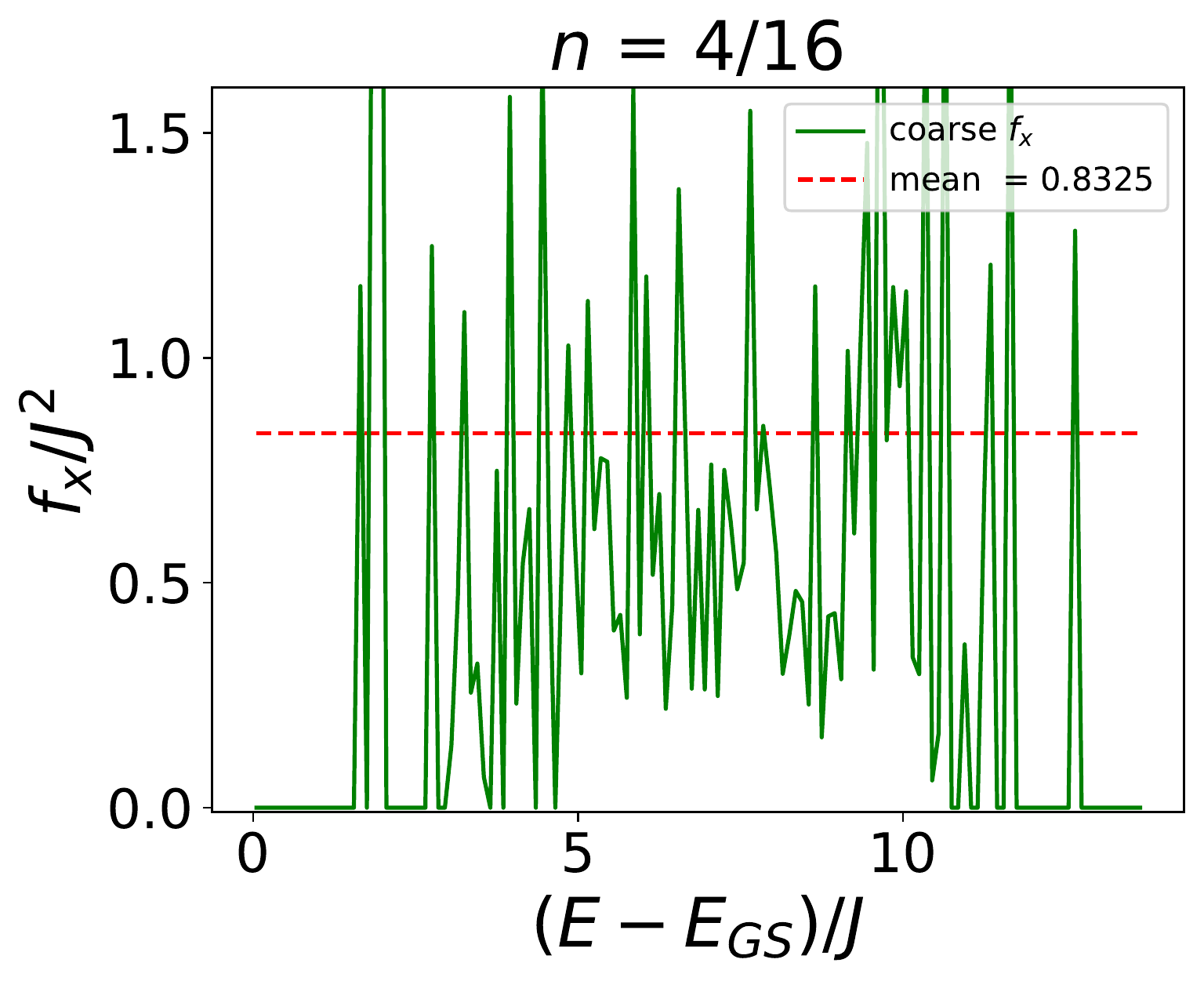}
\includegraphics[width=0.33\linewidth]{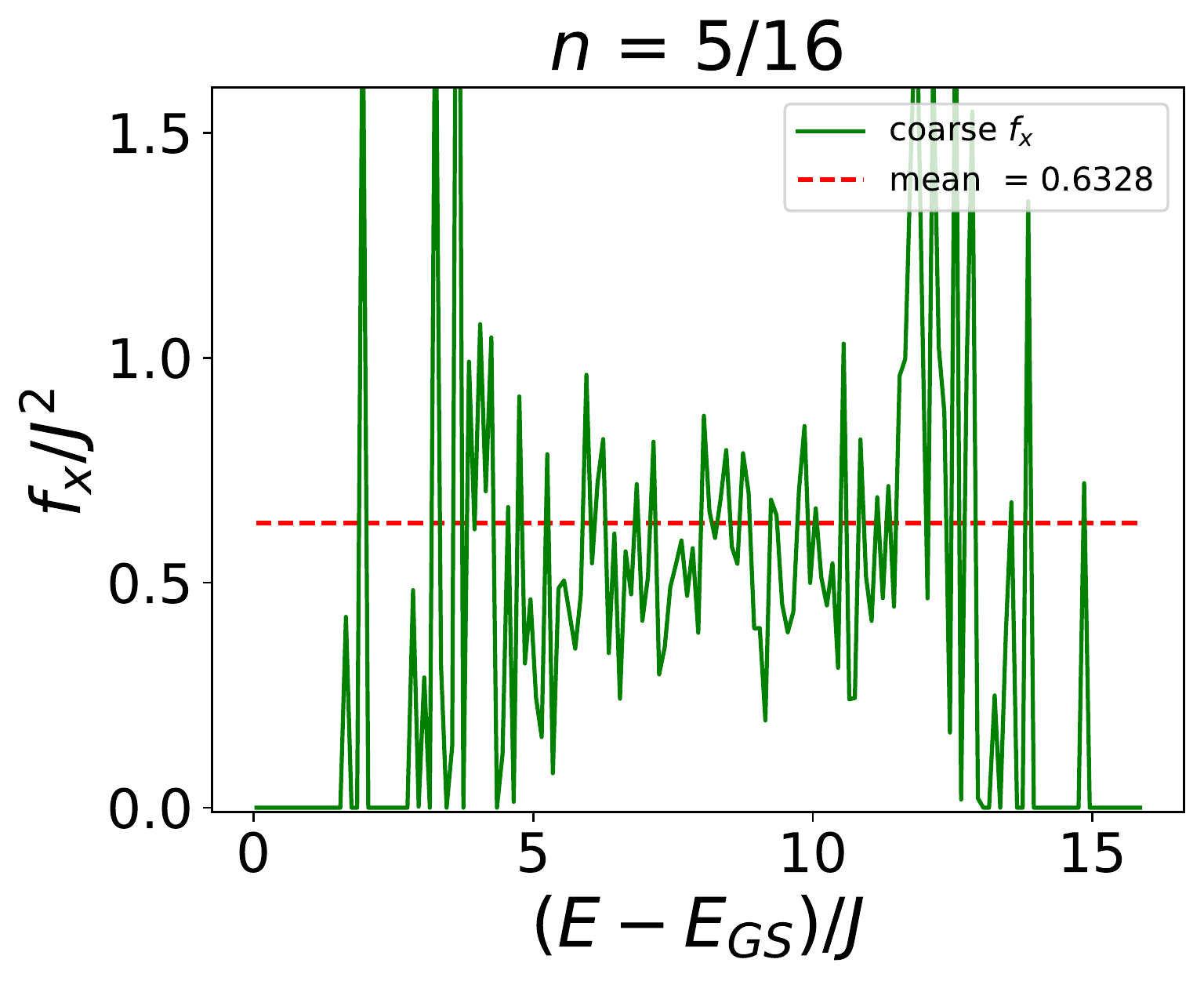}
\includegraphics[width=0.33\linewidth]{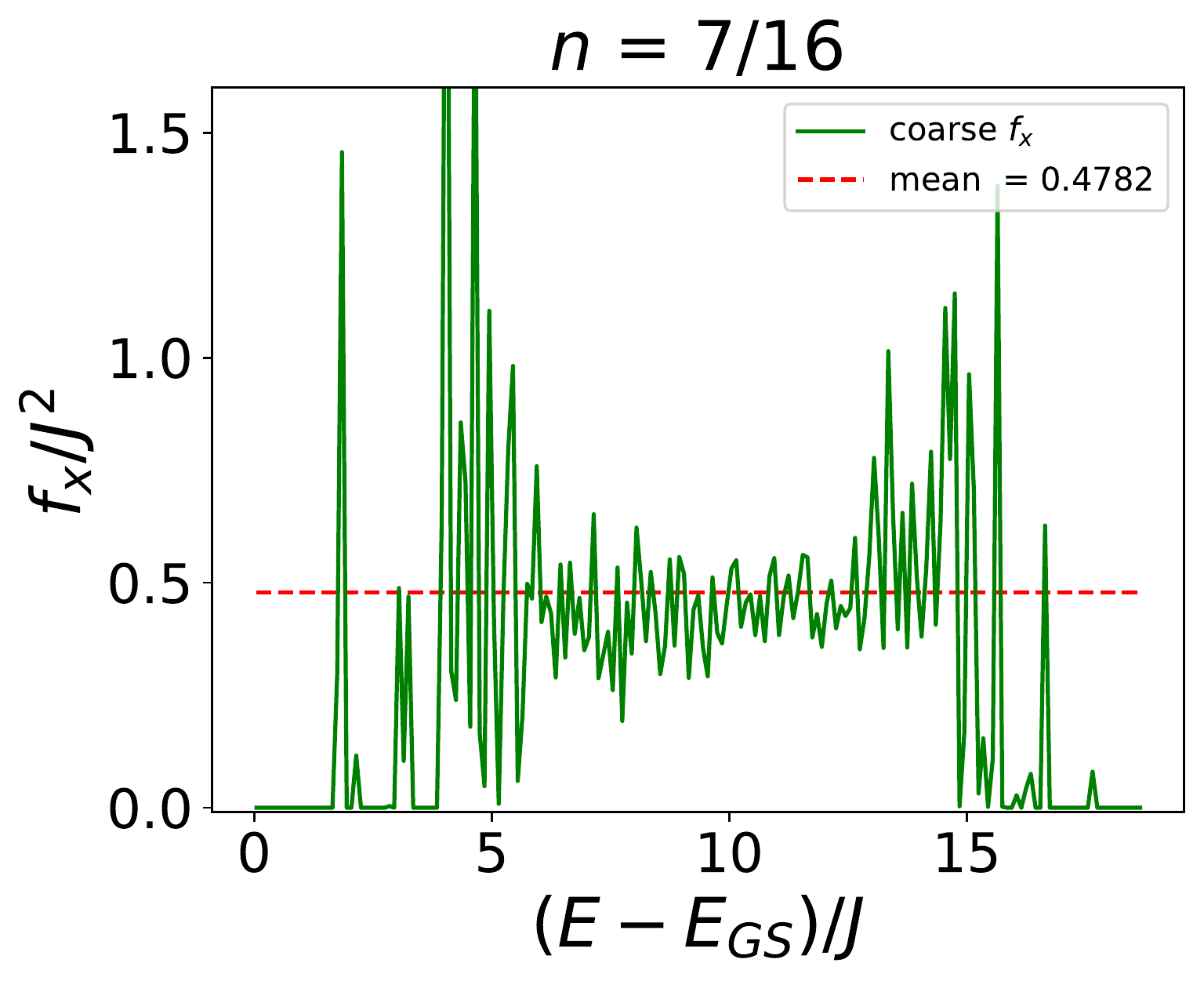}
\includegraphics[width=0.33\linewidth]{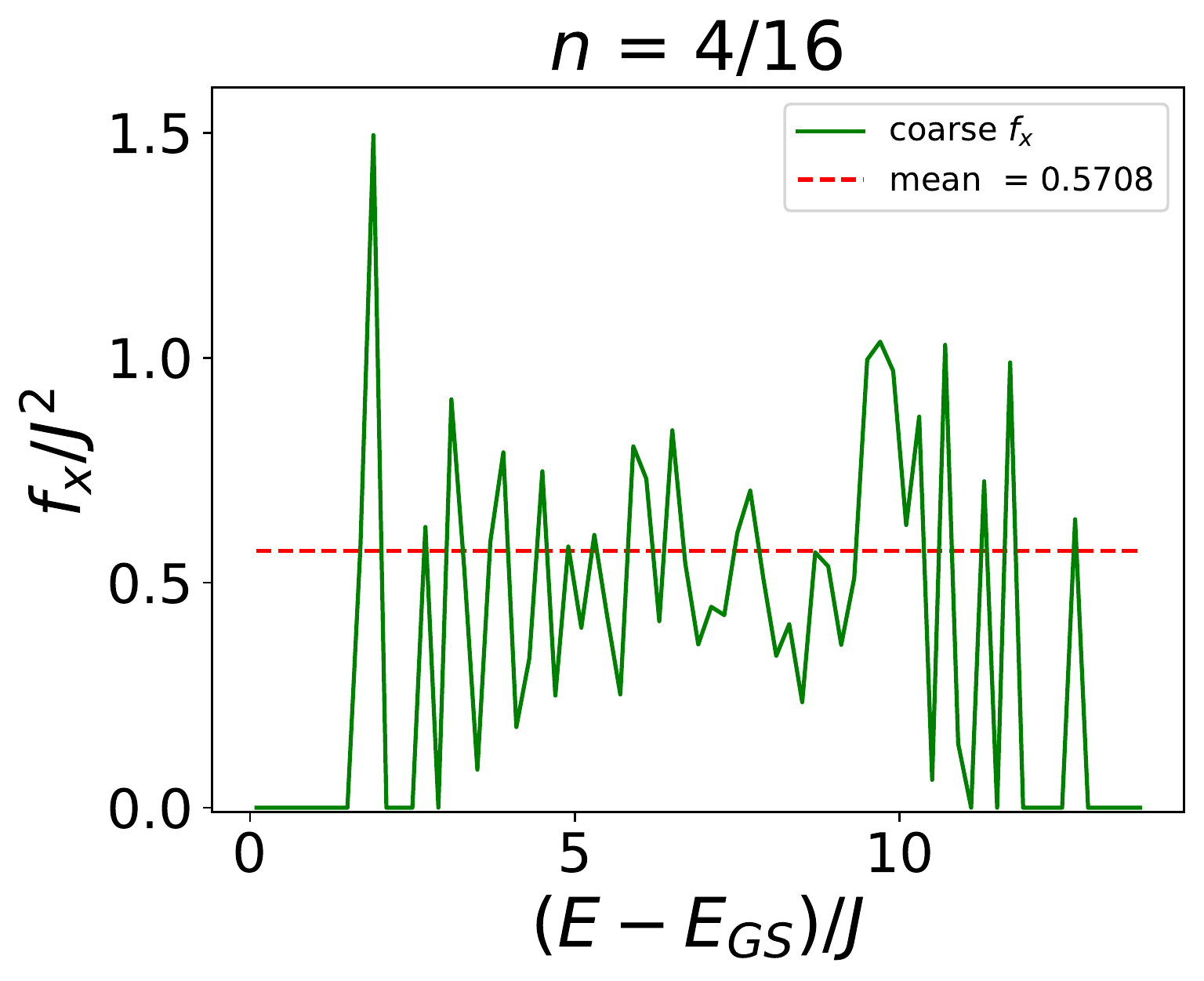}
\includegraphics[width=0.33\linewidth]{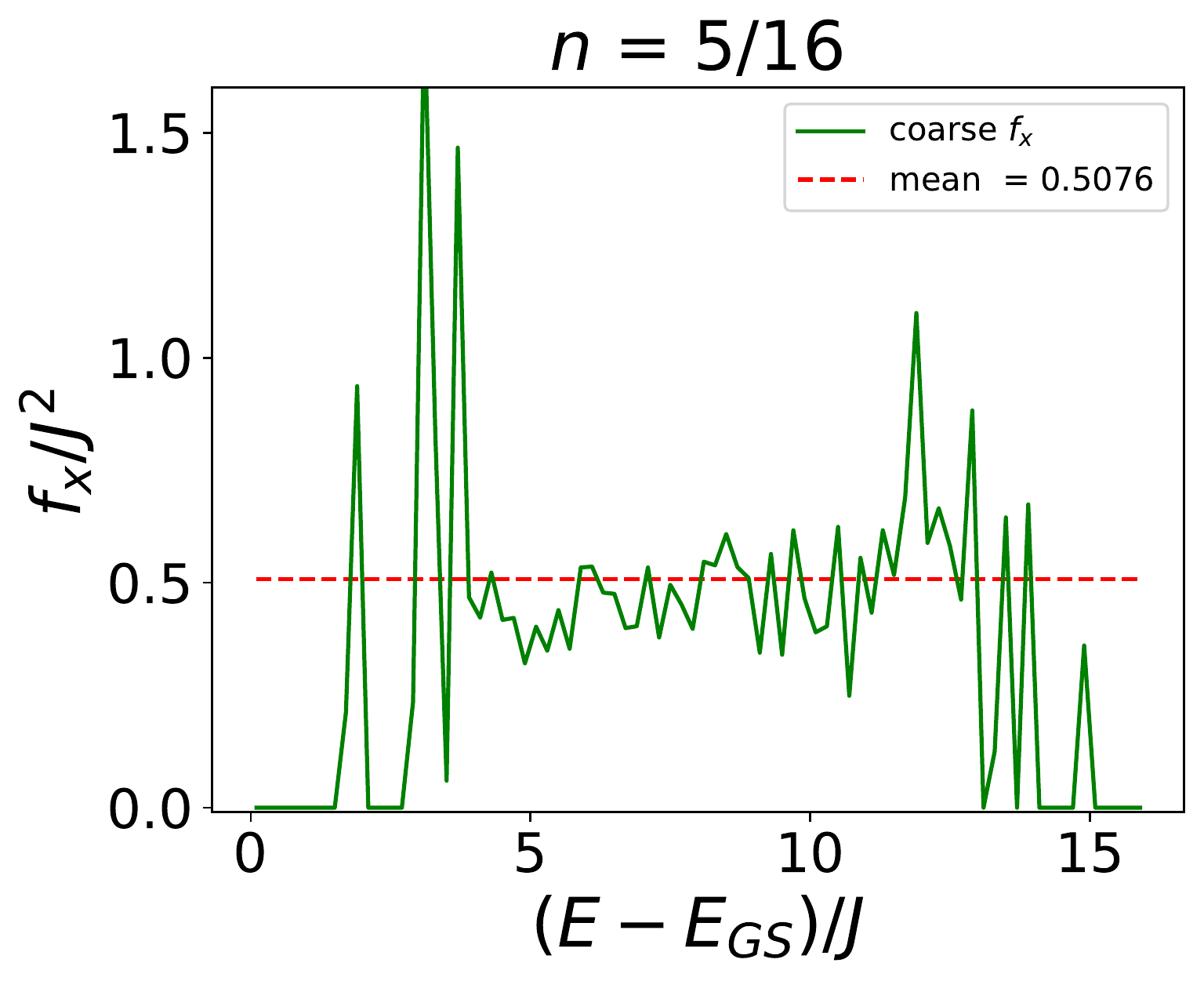}
\includegraphics[width=0.33\linewidth]{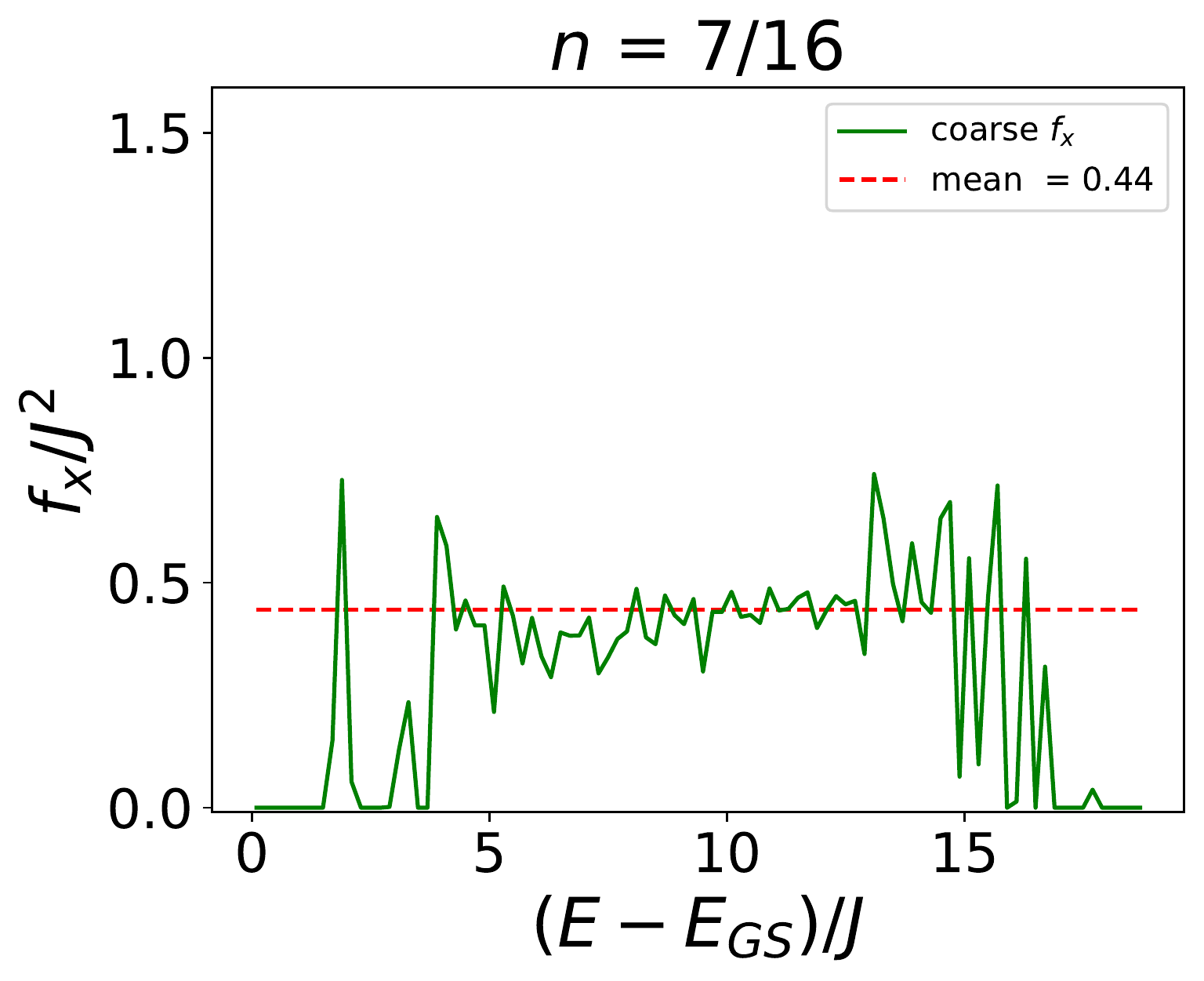}
\includegraphics[width=0.33\linewidth]{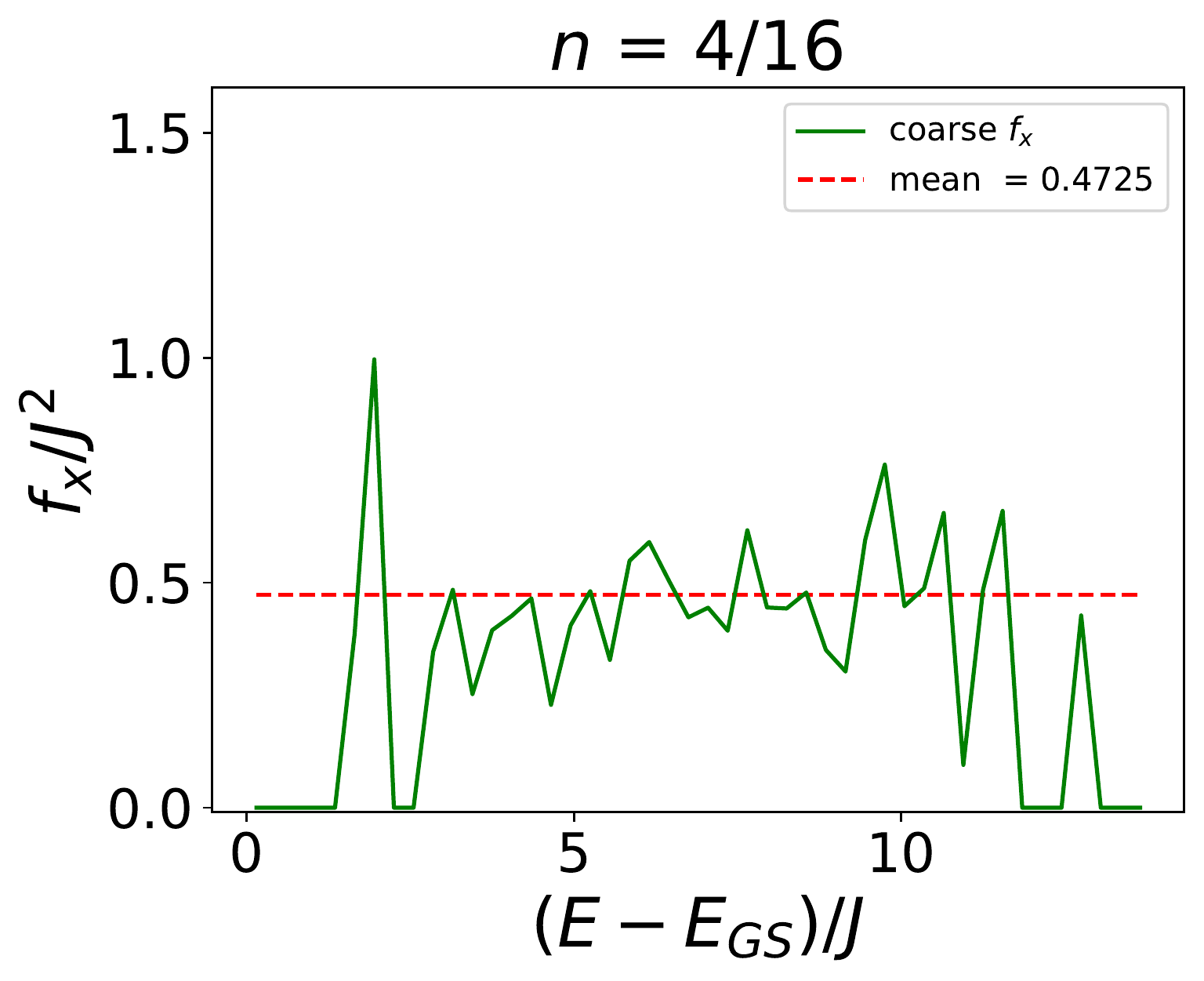}
\includegraphics[width=0.33\linewidth]{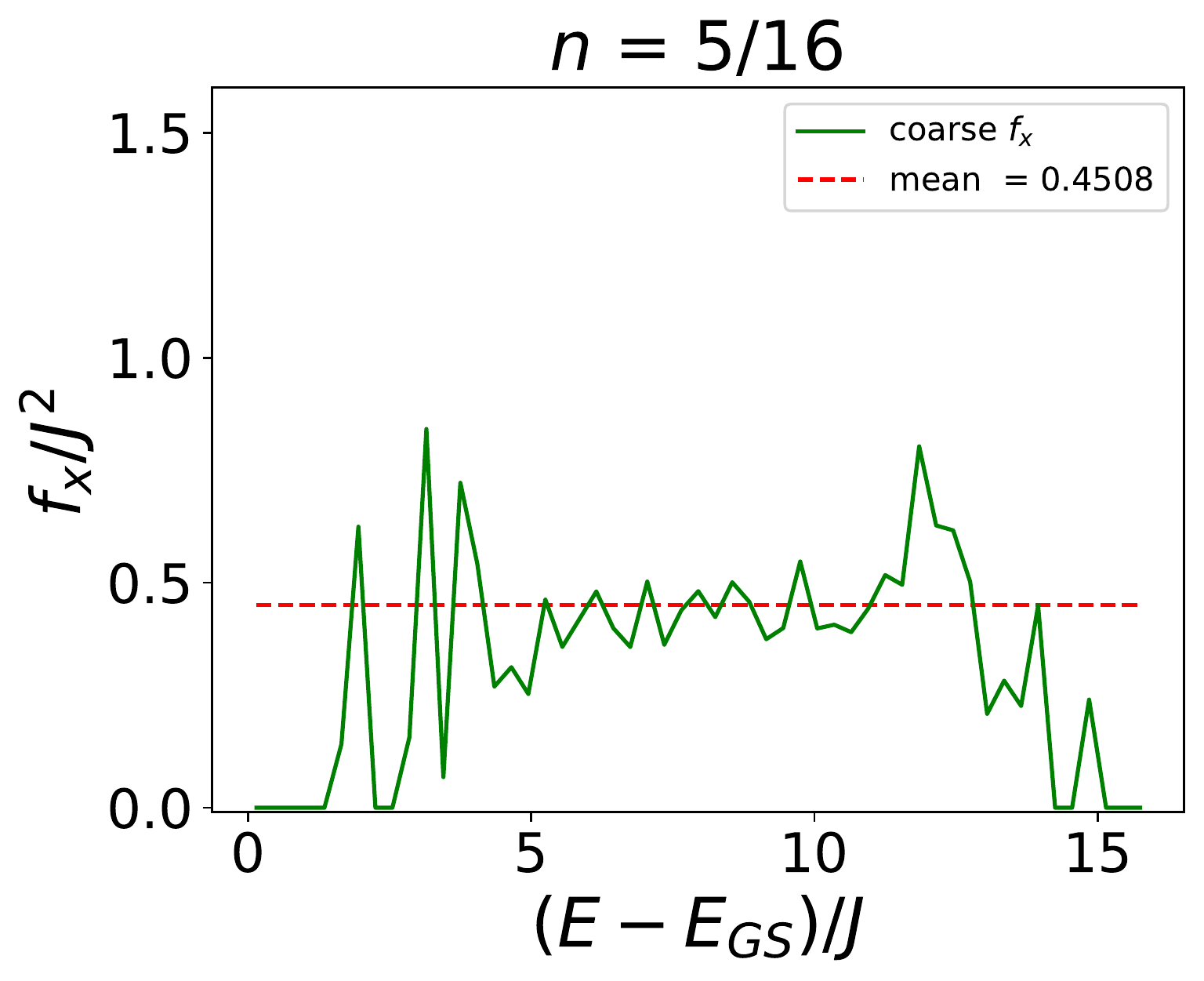}
\includegraphics[width=0.33\linewidth]{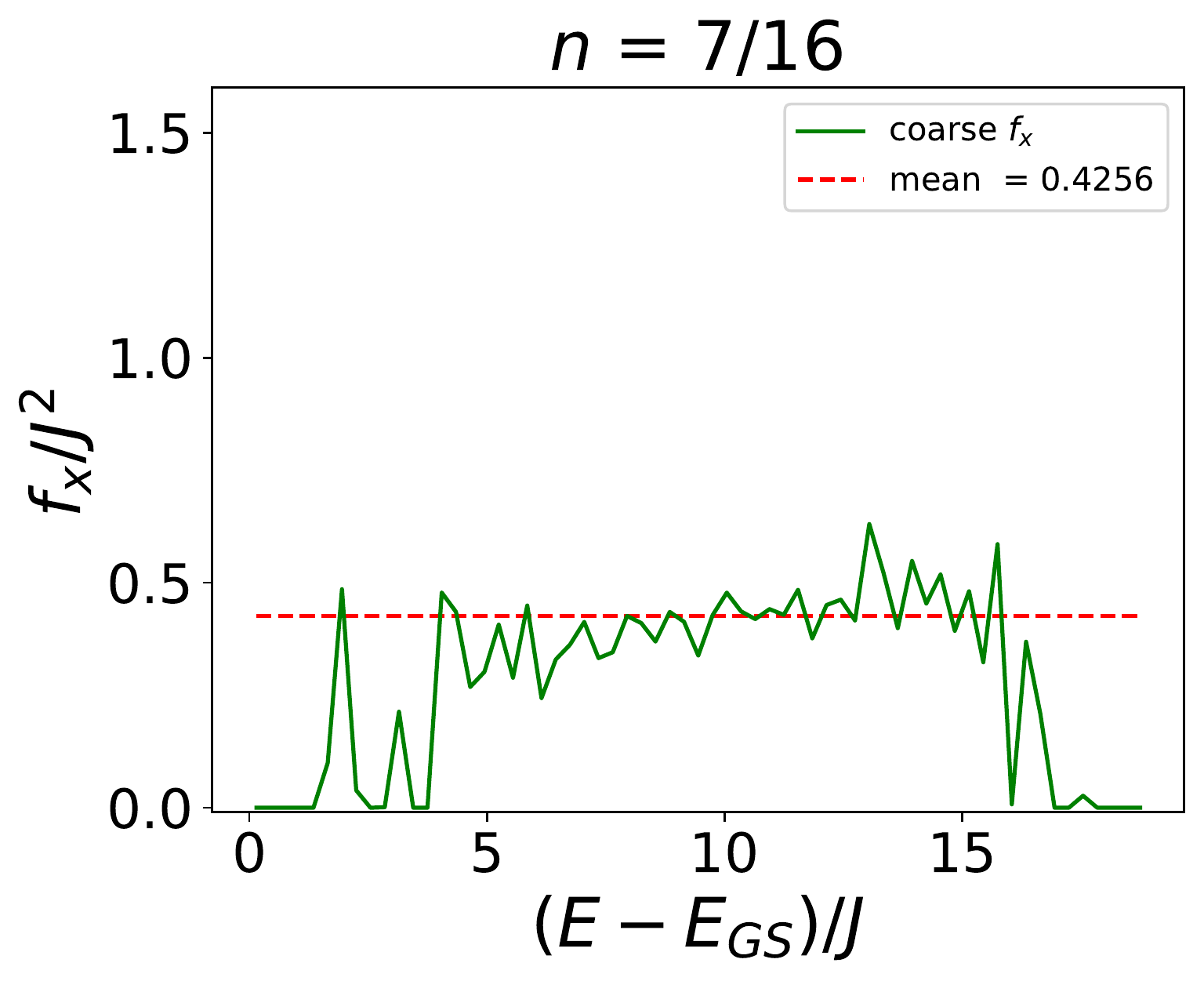}
\caption{$f_x(E)$ for the spin conductivity of the $4\times 4$ 2D square lattice nearest neighbor spin 1/2 Heisenberg model using the box binning protocol, with bin size $\eta$; $\eta=0.10 J$ (top row), $\eta=0.20 J$ (middle row) and $\eta = 0.30 J$ (bottom row). The representative magnetizations (equivalently fillings of spin-up particles in a sea of down spins, denoted by $n$) chosen are identical to the Lorentzian protocol. In each case, the ground state energy $E_{GS}$ has been subtracted out.
}
\label{fig:ffn_Heisenberg_box}
\end{figure}

\section{Broadening and binning for dynamical properties}

In the manuscript, we discussed the need for a Lorentzian broadening parameter $\eta$ to evaluate the conductivity, and the $f$-function. In Fig.~\ref{fig:ffn_spinless_Hubbard_1},~\ref{fig:ffn_spinless_Hubbard_2} and ~\ref{fig:ffn_Heisenberg_1} we present more data for the $f$-function for both the spinless Hubbard and the Heisenberg model, for different values of $\eta$. We also use this section to show results for other representative particle fillings or magnetizations not presented in the manuscript.

For the spinless Hubbard and Heisenberg models, the calculation proceeds in two steps. In the first step we evaluate $f_x(E_n,|n\rangle)$ using the Lorentzian approximation to the delta function. Each eigenstate thus has a $f$-value associated with it. If there is an exact degeneracy, we average the $f$-values for all states in the degenerate subspace and assign each state the same $f$-value. Once all the $f$-values are obtained, we plot their histogram which has been shown in the inset of our plots. This allows us to visually inspect how flat the $f$-function is. In the second step, we compute the coarse grained function by discretizing energy in units of $\eta$. The $f$-values of all eigenstates (computed in step 1) that lie in the energy window $(\gamma \eta, \gamma \eta +\eta)$ (where $\gamma=0,1,2...$ is an integer) are averaged to yield $f\Big(E= \Big(\gamma+ \frac{1}{2} \Big) \eta \Big)$. This coarse-grained function is plotted in the main panels of Fig. 1 of the text.

We have also explored a different coarse graining protocol, based on simple (box) binning of energy space. In this procedure all eigenstates are organized into bins of width $\eta$. The $f_{\alpha}(E)$ function is evaluated using,
\begin{equation}
	f_{\alpha}\Big(E = \Big(\gamma + \frac{1}{2} \Big) \eta \Big) = \frac{\pi}{ \eta \times \text{Number of states in bin} \; \gamma}\sum_{m,n \text{in same bin} \; \gamma} |I^{\alpha}_{nm}|^2
\label{eq:fboxed}
\end{equation}
where $\gamma=0,1,2...$ is an integer and represents the bin number for energies in the range $\gamma \eta$ to $(\gamma+1)\eta$.

Representative results of this binning protocol for the $4 \times 4$ 2D Heisenberg model are shown in Fig~\ref{fig:ffn_Heisenberg_box}. We observe that the $f$-function is indeed flat. However, for the same $\eta$ there is a difference between the box and Lorentzian protocols which gets smaller for large Hilbert spaces and small $\eta$. This is not too surprising, since the box protocol ignores inter-bin current contributions, an effect that becomes smaller for smaller bin widths $\eta$, and with larger number of eigenstates, both of which are required to reach the true continuum limit.

In either protocol, $\eta$ is chosen to be greater than the average level spacing. The average level spacing can be estimated from the lowest and highest many-body energy eigenstate and the size of the Hilbert space. We give some estimates here: For the spinless Hubbard model on the $4 \times 4$ cluster and $n=5/16$, the average level spacing for the case of $V/t=8$ is approximately $0.01 t$, which is a factor of 10 smaller than the value of the smallest $\eta$ chosen. For the Heisenberg model (hard-core boson model) at a (hard-core bosonic) filling of $7/16$, the average level spacing is about $0.002 J$, a factor of 50 smaller than the smallest $\eta$ chosen. 

To summarize our arguments: our results are applicable in the regime where $T>\eta$, and we have ensured $\eta >$ level spacing. It must be noted that even when there is a large gap between the low-energy manifold and the liquid states (as is the case for large $V/t$ in the spinless Hubbard model),  $\eta$ is still larger than level spacing in each of the manifolds individually. Since we have used values of $\eta$ ranging from $0.1t$ to $0.3t$, our results for the resistivity should be valid for $T>0.1t~\mathrm{to}~0.3t$, and this is the regime for which we present our results.

\section{Additional details on the projected Hubbard model}

\begin{figure}
\includegraphics[width=0.5\linewidth]{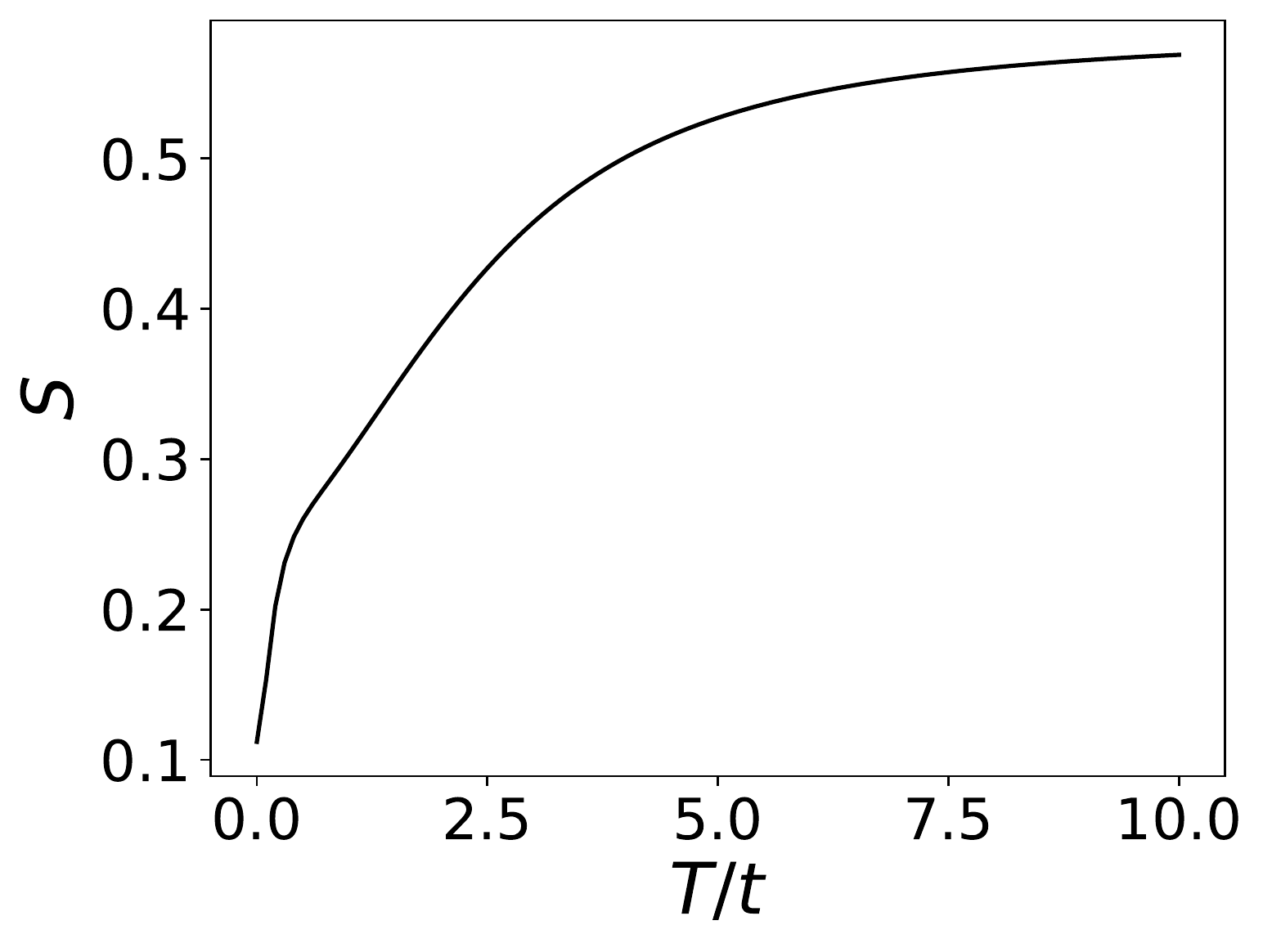}
\caption{Temperature dependence of the entropy density $S$, for the spinless Hubbard model with $V/t=4.3$ and filling $7/16$ ({\it i.e.} one hole away from half filling), after the low energy manifold of 16 states is projected out of the calculation. The remaining states form the ``incoherent quantum liquid" discussed in the main text, that is now shifted downwards in energy to low energies. Remarkably, this liquid displays an entropy density at low $T$ that is much lower than the combinatorial entropy density of the Hilbert space ($\ln (16!/(7!\times 9!) - 16)/16 \approx 0.584$), while the many-body states relevant to the low-$T$ regime still produce a $T$-linear resistivity with the same slope (and hence the same value of $f(E)$), as those relevant to the high-$T$ regime that posess the ``trivial" combinatorial entropy. Therefore, there is no temperature driven crossover in slope of the $T$-linear resistivity, while a corresponding temperature driven crossover does exist in the entropy density. Further investigation at larger system sizes is required to determine if $S(T\rightarrow0)$ goes to zero or takes on a non-zero value \cite{ParcolletGeorges,SachdevComplexSYK} in the thermodynamic limit.}
\label{fig:ST_spinless_Hubbard}
\end{figure}

The spinless nearest-neighbor Hubbard model close to half filling, with the low energy manifold projected out, provides a particularly appealing illustration of $f$-invariance, as the $f$-invariance holds across energy regimes that are different in terms of other physical properties. We show this as follows: in this model, since there are no spin fluctuations, entropy and specific heat are therefore directly related to the charge fluctuations only, which also control charge transport. Therefore, we can look at the entropy density to characterize different physical regimes relevant to charge transport. The entropy density $S$ at fixed particle density is given by the usual formula for the canonical ensemble
\begin{equation}
S = -\frac{1}{N_s}\frac{dF}{dT} = \frac{1}{N_s}\left(\ln Z + \frac{1}{Z T}\sum_n E_n e^{-\beta E_n}\right),  \end{equation}
where $F = - T\ln Z$ is the canonical Helmholtz free energy. We find that there is a “low entropy” regime at low $T$, with a crossover to a “high entropy” regime (which has the full combinatorial entropy of the Hilbert space) at high $T$ (Fig. \ref{fig:ST_spinless_Hubbard}). However, both regimes display $T$-linear resistivity with the {\it same} slope despite having different amounts of entropy (Fig. \ref{fig:projected_rho_spinless_Hubbard}).

\begin{figure}
\includegraphics[width=0.8\linewidth]{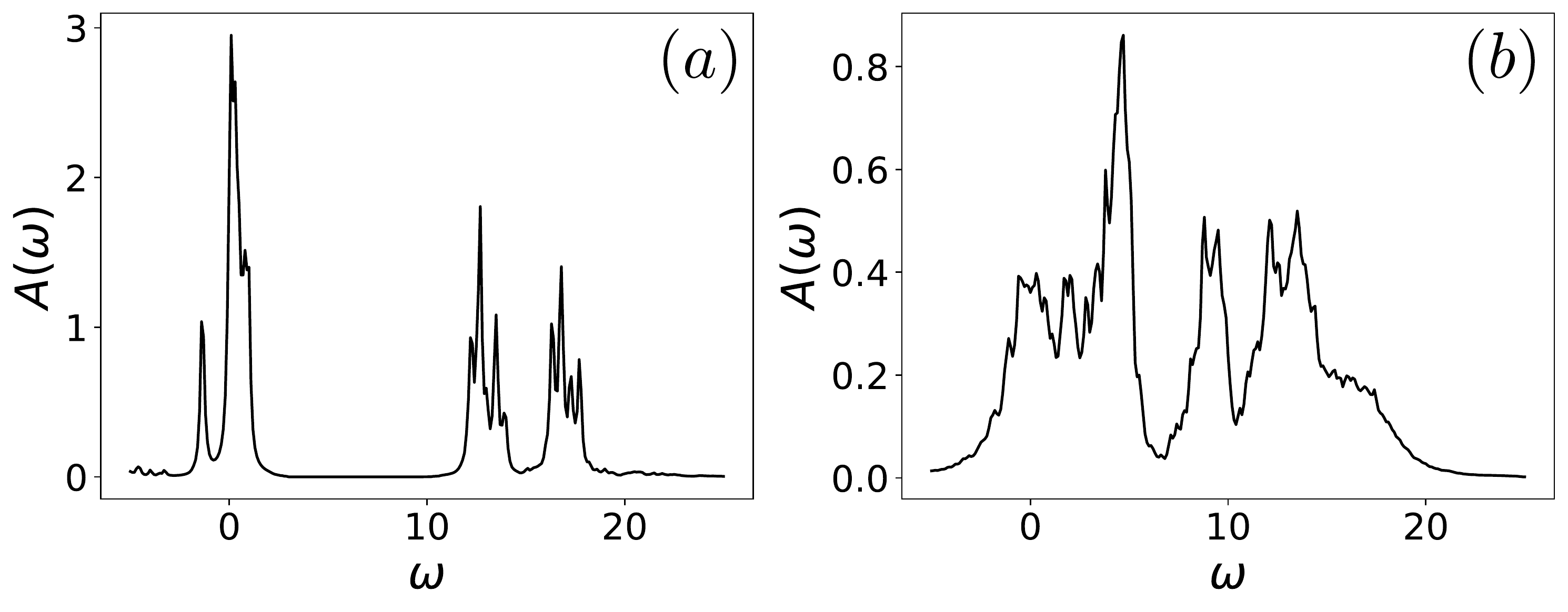}
\caption{(a) Single-particle spectral function $A(\omega)$, for the spinless Hubbard model with $V=4.3$, $t=1.0$, filling $7/16$, and temperature $T=1.0$. (b) Single-particle spectral function, after the low energy manifold of 16 states is projected out of the calculation. UV spectral weight from the upper Hubbard bands at $\omega \sim 3V$ and $\omega \sim 4V$ in (a), that represent energy costs of certain processes adding one particle to the system, is transferred down to lower frequencies by the projection procedure in (b), yielding a broad gapless continuum of spectral weight extending across energy scales. The redistribution of spectral weight down towards low energies, which arises from shifting UV states downward in energy, is ultimately what leads to $T$-linear resistivity existing at low $T$ as well in the projected model, with the same slope as the $T$-linear resistivity in the high $T$ regime. The spectral functions in (a) and (b) are both computed using their spectral decompositions, with a Lorentzian broadening $\eta=0.1$ of the $\delta$-functions.}
\label{fig:SF_spinless_Hubbard}
\end{figure}

We also computed the single-particle spectral function $A(\omega) = -2\mathrm{Im}[G^R(\omega)]$, where $G^R(t) = i\theta(t)\mathrm{Tr}[e^{-\beta H}\{c_j(t),c_j^\dagger(0)\}]$, and $G^R(\omega)=\int_{-\infty}^{\infty} dt~G^R(t)e^{i\omega t}$, for this model. The results are shown in Fig. \ref{fig:SF_spinless_Hubbard}. Without projecting out the low energy manifold, the spectral function has some weight at $\omega\sim 0$. This feature arises from the addition or removal of one particle with (almost) no change in energy; when the added or removed particle does not have any immediate neighbors, 
there is no $V$ energy cost associated with it. The spectral function also shows weight at $\omega\sim 3V$ and $\omega \sim 4V$; these features arise from the addition of one particle, and hence the introduction of additional $V$ costs. Upon projecting the low energy manifold out, the high frequency spectral weight is redistributed down to lower frequencies as a gapless continuum of states. The resulting UV-IR mixing causes low energy states to be similar to high energy ones, which also has the effect of extending the $T$-linear resistivity of high temperatures down to low $T$ with the same slope, and out of the bad metallic regime.  

\section{Details of computation of $f(E)$ for SYK}

For the SYK calculations, the large number of eigenstates on both dots combined makes the implementation of the full Lorentzian no longer computationally feasible. Instead, we used the box binning method mentioned above, and computed the coarse grained function directly without evaluating $f(E_n,|n\rangle)$.

For each realization of the random couplings $J_{ijkl}$, we defined a grid of energy points $0,\Delta E,~...~,E_{\mathrm{max}}-\Delta E,E_{\mathrm{max}}$, where $E_{\mathrm{max}}$ is the many-body bandwidth of the two dot system, and the ground state energy $E_{\mathrm{GS}}$ is subtracted out so that the smallest energy is zero. We took $\Delta E \approx E_{\mathrm{max}}/50$. Then, for each $E$ point, we defined a bin centered at it of width $\eta$, that is the smaller of $\Delta E$ and the width required for the bin to contain approximately $500$ states. We performed a binary search to determine the appropriate value of $\eta$ for each value of $E$, and then computed $f(E)$ according to Eq. (\ref{eq:fboxed}). Finally we averaged the values of $f(E)$ at these grid points over several realizations of $J_{ijkl}$.

While diagonalizing each of the two SYK dots, we ignore the states that have $0$, $1$, $N-1$, or $N$ particles on a dot. For a total of $N$ particles on the two dots (half filling), these states are all at zero energy, and are irrelevant in the large $N$ limit, as they then form a vanishingly small subset of the total Hilbert space. However, at small $N\le 10$, they do create a visible artifact in $f(E)$ in the middle of the many-body band, which vanishes as $N$ is increased, as it should. By ignoring these states, this distracting artifact is removed from the plots of $f(E)$ even for small $N$.

Finally, we remark on how we obtained the large $N$ results for $f(E)$ in SYK in Fig. {\color{magenta} 4} of the main text. Analytic calculations \cite{Cha_2020} have shown, in the large $N$ limit, that $T\sigma(0,T) = Nt^2\sqrt{\pi}/2$ for $T\ll J$, and $T\sigma(0,T) \approx 1.05Nt^2\sqrt{\pi}/2$ for $J\ll T\ll NJ$. Since $f$-invariance is required for $T$-linear resistivity over extended energy regimes, it follows that we can substitute the values of $T\sigma(0,T)$ (or $\rho(T)/T$) for $f(E)$ in the corresponding energy regimes. Remarkably, our numerics suggest that $f$-invariance continues to hold as one enters the even higher energy regime of $E \sim NJ$, {\it i.e.} the middle of the many-body band.

\end{document}